\documentclass{aa}  

\usepackage{graphicx}
\usepackage{txfonts}
\usepackage{xcolor}
\usepackage[colorlinks=true,linkcolor=blue,citecolor=blue]{hyperref}
\usepackage[normalem]{ulem}

\begin{document}

   \title{Electron-proton co-acceleration on relativistic shocks\\ in extreme-TeV blazars}

   \author{Andreas Zech
          \inst{1}
          \and
          Martin Lemoine
          \inst{2}
          }

   \institute{Laboratoire Univers et Théories, Observatoire de Paris, Université PSL, CNRS, Université de Paris, 92190 Meudon, France\\
              \email{andreas.zech@obspm.fr}
         \and
            Institut d'Astrophysique de Paris, CNRS -- Sorbonne Universit\'e, 98 bis boulevard Arago, 75014 Paris, France\\
             \email{lemoine@iap.fr}
             }

   \date{}

 
  \abstract
   {}
   {The multi-wavelength emission from a newly identified population of `extreme-TeV' blazars, with Compton peak frequencies around 1\,TeV, is difficult to interpret with standard one-zone emission models. Large values of the minimum electron Lorentz factor and quite low magnetisation values seem to be required.
   }
   {We propose a scenario where protons and electrons are co-accelerated on internal or recollimation shocks inside the relativistic jet. In this situation, energy is transferred from the protons to the electrons in the shock transition layer, leading naturally to a high minimum Lorentz factor for the latter. A low magnetisation favours the acceleration of particles in relativistic shocks.}
   {The shock co-acceleration scenario provides additional constraints on the set of parameters of a standard one-zone lepto-hadronic emission model, reducing its degeneracy. Values of the magnetic field strength of a few mG and minimum electron Lorentz factors of $10^3$ to $10^4$, required to provide a satisfactory description of the observed spectral energy distributions of extreme blazars, result here from first principles. While acceleration on a single standing shock is sufficient to reproduce the emission of most of the extreme-TeV  sources we have examined, re-acceleration on a second shock appears needed for those objects with the hardest $\gamma$-ray spectra. Emission from the accelerated proton population, with the same number density as the electrons but in a lower range of Lorentz factors, is strongly suppressed. Satisfactory self-consistent representations were found for the most prominent representatives of this new blazar class.}
   {}


   \maketitle
%
\section{Introduction} 
\label{sec:intro}
Blazars -- {\it } active galactic nuclei (AGN) with a jetted relativistic outflow pointing towards us -- are commonly classified according to their characteristic double-humped spectral energy distribution \citep[SED;][]{1995PASP..107..803U,1998MNRAS.299..433F}. At the tip of the so-called blazar sequence \citep{1995ApJ...444..567P,2017MNRAS.469..255G}, extreme blazars form a new population of BL Lacs \citep{2001A&A...371..512C}, designated as `ultra-high-frequency-peaked BL Lac objects' (UHBLs) given the very high frequencies of their two emission peaks \citep{Biteau2020}. The sub-class of `extreme-TeV' sources presents rather unusual properties, with the higher energy peak beyond $\sim 1\,$TeV and a hard spectrum in the $\gamma$-ray range. These objects still defy a  theoretical interpretation in terms of the commonly used one-zone radiative models, which are generally successful in the representation of SEDs from BL Lac objects. 

Although ten or so extreme-TeV sources are known to date \citep{Biteau2020}, current estimates based on data from the {\it Fermi}-LAT $\gamma$-ray telescope predict about five times more candidate sources that need to be confirmed with future instruments in the TeV range \citep{Costamante2020}. For five of the known extreme-TeV blazars, largely simultaneous SEDs with a good multi-wavelength coverage from the infrared to the TeV band have become available, and they have already provided important clues on the physics of dissipation in these sources \citep{Costamante2018}. 
 
Arriving at a satisfactory description of extreme-TeV sources in the standard one-zone leptonic framework seemingly requires two essential ingredients \citep[e.g.][]{1998ApJ...509..608T,2002A&A...386..833G,2006MNRAS.368L..52K,Costamante2018}: (1) lower magnetic fields ($\lesssim 10\,$mG) than for more common high-frequency-peaked BL Lac objects (HBLs), corresponding to magnetic field energy densities well below equipartition, and (2) a peculiar lepton distribution with most of the energy carried by particles of a large Lorentz factor, $\overline\gamma_e \sim 10^3 - 10^4$. The former requirement avoids an excessive softening of the $\gamma$-ray spectrum by synchrotron cooling and is necessary to reproduce the large separation between the synchrotron and synchrotron self-Compton (SSC) peak frequencies, while the latter is needed to accommodate the steep $\gamma$-ray spectrum and high peak frequency. Alternative models have been devised \citep[e.g.][]{2000NewA....5..377A,2008ApJ...679L...9B,2011ApJ...740...64L,2014MNRAS.438.3255T,Cerruti2015,2017MNRAS.466.3544C,2020MNRAS.491.2198T}, but they all come at the price of greater complexity and most often invoke other extreme parameter values, which often lack experimental or physical justification. For instance, lepto-hadronic models can provide reasonable SEDs with less extreme values of $\overline\gamma_e$, but at the expense of a high jet power, mostly carried by the accompanying population of ultra-relativistic protons or the magnetic field \citep[e.g.][]{Cerruti2015}.

In the present study we take at face value the low magnetisation inferred from the SEDs of extreme blazars and propose revisiting the one-zone model, adopting well-motivated microphysical prescriptions for particle acceleration, under the assumption that electrons are co-accelerated with protons in relativistic internal or recollimation shocks. Two key motivations of our study are the following: (1) at low magnetisation, relativistic shock acceleration is expected to dominate the physics of dissipation and of the generation of non-thermal power laws, and (2) electrons can be efficiently preheated in the transition layer of electron-ion relativistic shocks, up to a fraction of the thermal energy of shocked ions, implying large minimum Lorentz factors for the electron population, as seemingly required by phenomenology~\cite[e.g.][and references therein]{2020Galax...8...33V}. We further aim at making our approach as self-contained as possible, by fixing the microphysical parameters describing the acceleration process from  first principle studies of relativistic shock models. Our main objective is to provide a critical discussion of such models to reproduce the SEDs of extreme blazars, in a modern context. This contrasts with generic one-zone models, which instead parameterise the accelerated particle population and adjust the parameters to the observed SEDs. 

We organise the discussion as follows. In Sect. \ref{sec:coacceleration} we discuss existing  constraints on the microphysical parameters of shock acceleration. The standard one-zone SSC model is then extended to include a population of protons, and physical constraints from the considered acceleration mechanism are translated into constraints on the parameters that describe the two particle populations. Section~\ref{sec:application} then applies the resulting model to the set of well-covered SEDs. In a first scenario, generic solutions with a minimum number of free parameters are presented. A second scenario applies realistic parameters for acceleration on a mildly relativistic internal or recollimation shock, while a third scenario accounts for the possibility of re-acceleration at multiple shock waves, which appears necessary to reproduce the hard $\gamma$-ray spectra of certain objects. A critical discussion of this model, as well as the direct consequences for our understanding of the jets of extreme blazars, is given in Sect.~\ref{sec:discussion}.


\section{A theoretical model for e-p co-acceleration in blazar internal shocks}
\label{sec:coacceleration}

\subsection{A large $\overline\gamma_e$}\label{sec:genrem}
We define the Lorentz factor $\overline\gamma_e$ as the average over the population,
\begin{equation}
    \overline\gamma_e = \frac{1}{n_e}\int{\rm d}\gamma_e\,\gamma_e\,\frac{{\rm d}n_e}{{\rm d}\gamma_e}\,,
    \label{eq:ovgamma}
\end{equation}
with $n_e=\int{\rm d}\gamma_e\,{\rm d}n_e/{\rm d}\gamma_e$ the electron number density. Blazar SEDs are generically well represented by radiative leptonic emission from a (broken) power-law population with spectral indices $s_1$ (resp. $s_2$) below (resp. above) a break Lorentz factor $\gamma_{e,\rm b}$. For such a distribution of Lorentz factors, $\overline\gamma_e \sim\,\gamma_{e,\rm min}\,\left(\gamma_{e,\rm b}/\gamma_{e,\rm min}\right)^{2-s_1}$ if $s_1<2<s_2$, and $\overline\gamma_e\sim\,\gamma_{e,\rm min}$ if $2<s_1<s_2$. The energy density $u_e$ of the population can then be written $u_e \sim \overline\gamma_e \,n_e\, m_e c^2$. 

Therefore, to convert a population of electrons (or pairs) of initial Lorentz factor $\gamma_0$ to the above power-law spectrum necessitates the dissipation of an energy reservoir that is $\overline\gamma_e/\gamma_0$ times larger than the initial particle energy. In this sense, the large inferred value of $\overline\gamma_e$ provides an important clue regarding the amount of energy that is dissipated and its origin, given that, in the models of \citet[]{Costamante2018}, $\overline\gamma_e \sim 10^3 - 10^4$ arises as a generic prediction independently of the value of $s_1$. 

In a (co-moving) magnetic field of strength $B=10\,B_{-2}\,$mG, electrons of Lorentz factor $\gamma_e=10^4\,\gamma_{e\,4}$ cool through synchrotron radiation on a length scale 
\begin{equation}
l_{\rm c}\simeq 8\Gamma_{\rm j}\, B_{-2}^{-2} \, \gamma_{e\,4}^{-1}\, \rm pc\,,
\label{eq:cooling_length}
\end{equation}
as measured in the source rest frame, and expressed in terms of $\Gamma_{\rm j}$ the jet Lorentz factor. Hence, one cannot {\it a priori} exclude that electrons gained their large average Lorentz factors in some primary dissipation region, located well upstream of the acceleration region where the power-law population is generated. Such a scenario is nevertheless constrained by the apparent lack of radiation associated with that putative primary dissipation process, as well as by the shortened cooling length in the primary dissipation region with enhanced magnetic strength. In the absence of a definite model for the evolution of the magnetic field strength, electron distribution etc. along the jet, it is difficult to make this statement more quantitative as the synchrotron luminosity scales with the combination $\Gamma_{\rm j}^4n_e R^3 u_B \gamma_{e,\rm min}^{s-1}\gamma_{e,\rm max}^{3-s}$ for a single power law of index $s<3$ extending from $\gamma_{e,\rm min}$ to $\gamma_{e,\rm max}$, where $u_B$ denotes the magnetic energy density, $n_e$ the electron density, $R$ the jet radius, and $\Gamma_{\rm j}$ its Lorentz factor. We thus leave this possibility open in the following.

In this context, it is particularly interesting to investigate the alternative possibility that the heating of electrons to large minimal Lorentz factors is, in fact, associated with the acceleration process itself. In the following, we focus on the case of shock acceleration, while alternative scenarios involving reconnection or turbulent acceleration are examined in Appendix~\ref{app:reconnection}.

\subsection{Electron heating in relativistic shocks}
\label{sec:shockheating}
Consider a shock front, moving at Lorentz factor $\gamma_{\rm sh}$ (and velocity $\beta_{\rm sh}c$) with respect to the un-shocked plasma. In the context of extreme blazars, we are mostly interested in the case of mildly relativistic or truly relativistic shocks ({\it i.e.} $u_{\rm sh}\equiv\gamma_{\rm sh}\beta_{\rm sh}\gtrsim 1$). We also assume here a shock of normal incidence, meaning that the upstream flow enters the shock front along the shock normal, whose direction is perpendicular to the shock front. We generalise this to the case of oblique shocks in Sect.~\ref{sec:oblique}.

If the composition is purely leptonic, the electrons are energised through shock crossing according to the standard Rankine-Hugoniot conditions, or rather their relativistic generalisation~\citep[e.g.][and references therein]{1999JPhG...25R.163K}. For a strong, sub-relativistic pair shock, the post-shock temperature reads $k_B T_e/m_e c^2 \simeq 0.19 u_{\rm sh}^2$, while for a strong, relativistic pair shock, $k_B T_e/m_e c^2 \simeq 0.24 u_{\rm sh}$. For reference, $k_B T_e/m_e c^2\simeq 0.7$ for $u_{\rm sh}=3$ in the mildly relativistic regime, close to the fully relativistic limit.

This situation changes rather dramatically in an electron-ion shock, assuming for simplicity an equal number of ions and electrons. Most of the energy that enters the shock front is now carried by the ions, but a fraction of it is given to the electrons, which are thereby preheated up to a fraction of equipartition.  More specifically, in the reference frame in which the shock front lies at rest, the kinetic energy density in (cold) protons is $e_p=\gamma_{\rm sh}\left(\gamma_{\rm sh}-1\right)n_p m_p c^2$ ($n_p$ is understood as a proper density here), corresponding to an energy-per-particle $\langle E_p\rangle = \left(\gamma_{\rm sh}-1\right)m_p c^2$, and similarly for the electrons with ${}_p\leftrightarrow {}_e$.  Efficient preheating of the electrons means that $\langle E_e\rangle $ becomes a substantial fraction of $\langle E_p\rangle$, instead of $\langle E_e\rangle / \langle E_p\rangle = m_e/m_p\ll1$, which would be expected if both species were to satisfy the fluid shock crossing conditions independently from each other.

The mechanism(s) through which electrons gain energy at the expense of ions in shock fronts remain somewhat elusive. In the relativistic, weakly magnetised limit, one promising scenario is that in which electrons undergo collisionless Joule heating, driven by a longitudinal electric field and by the effective gravity that results from the slowdown of the microturbulence that is self-generated in the shock precursor  \citep[see e.g.][]{2019PhRvL.123c5101L,Vanthieghem+21}. Phenomenological models of gamma-ray burst afterglows provide a nice empirical confirmation of this energy transfer between ions and electrons, as the electron energy fraction is almost always found to be within a factor of a few from the ion energy fraction \citep[e.g.][and references therein]{2015PhR...561....1K}.

In the absence of a definitive model, we can rely on particle-in-cell (PIC) simulations to provide estimates for the average electron Lorentz factor. Generally speaking, the preheating appears more efficient at lower magnetisations $\sigma$~\citep{2013ApJ...771...54S}, likely because of a larger precursor and because of the presence of intense Weibel-like microturbulence. For reference, we define $\sigma$ as the ratio of the magnetic energy density to the particle energy density in the rest frame of the un-shocked (ambient) plasma. Simulations also report a more efficient preheating in the fully relativistic regime -- meaning $\gamma_{\rm sh}\gtrsim 3-5$ -- than in the mildly relativistic regime $\gamma_{\rm sh} \sim 1-3$. For example, \cite{2019MNRAS.485.5105C} measure $T_e\simeq 0.3T_p$ and $k_B T_p\simeq 0.2m_p c^2$ for a mildly relativistic sub-luminal shock of velocity $\beta_{\rm sh}=0.8$, corresponding to $\gamma_{\rm sh}=1.7$, at magnetisation $\sigma \sim 0.007$ (here defined downstream). This electronic temperature corresponds to a mean Lorentz factor $\overline\gamma_e\sim 300$. At larger shock velocities, and smaller magnetisations, \cite{2011ApJ...726...75S} and \cite{2013ApJ...771...54S}  find that the fraction of energy stored in electrons increases up to half of that in the ions, implying $k_B T_e \sim 0.1 \gamma_{\rm sh} m_p c^2$, or a mean Lorentz factor $\overline\gamma_e\sim 600 \gamma_{\rm sh}$. 

In the sub-relativistic limit, and more generally in cases in which electron preheating is not efficient, the electrons suffer from a well-known injection problem into the Fermi acceleration cycle. In such cases, the energy density of the suprathermal tail of accelerated electrons represents a modest fraction $\epsilon_e$ of the energy that is incoming into the shock, for example $\epsilon_e\simeq 0.001$ for the case with $\beta_{\rm sh}=0.8$ discussed in \cite{2019MNRAS.485.5105C}. Correspondingly, the minimum Lorentz factor of the suprathermal electron power law -- that is, the Lorentz factor at which the power law emerges out of the thermal Maxwellian -- is substantially larger than $\overline\gamma_e$. In the relativistic limit, $\epsilon_e \gtrsim 0.1$ because preheating, hence injection is efficient and the minimum Lorentz factor of the accelerated power law is a factor of a few larger than $\overline\gamma_e$, at most. 

If electron preheating were inefficient, the electron distribution would be dominated by the thermal bump, hence it would not account satisfactorily for the observed SEDs of extreme blazars, which not only require a large $\overline\gamma_e$, but also a hard spectrum with index $s\sim 2$. In the following, we thus adopt $\epsilon_e = 0.1$ and $\overline\gamma_e\sim 600 \gamma_{\rm sh}$ as fiducial values, corresponding to relativistic, weakly magnetised electron-ion shocks. We also assume one electron per ion, as larger electron multiplicities are expected to lead to lower mean energy per particle, although this case has not been explicitly probed by numerical experiments.

\subsection{Fixing parameters from the microphysics of shock acceleration}
\label{sec:microphysics}
In the fully relativistic regime, meaning a shock Lorentz factor $\gamma_{\rm sh}\gtrsim 3-5$, particle acceleration appears restricted to the weakly magnetised limit, $\sigma\lesssim 10^{-3}$ (see {\it e.g.}~\citealt{2020Galax...8...33V} and references therein). Strictly speaking, this result applies to super-luminal shocks, but highly relativistic shocks are generically super-luminal as a consequence of the Lorentz boost (a factor $\gamma_{\rm sh}$) of the perpendicular magnetic field components in the shock rest frame. In the mildly relativistic limit, however, such effects disappear and sub-luminal configurations become about as likely. Consequently, particle acceleration can become efficient, even at relatively high magnetisations (see e.g. \citealt{2019MNRAS.485.5105C})

In the regime of low magnetisation that we are interested in, the size of the precursor is governed by the scattering of particles in the electromagnetic micro-turbulence that they themselves excite through beam-plasma instabilities~\cite[e.g.][]{2018MNRAS.477.5238P}, and which is responsible for the shock transition itself. This general problem has been studied in some detail in the limit of negligible initial magnetisation~\cite[e.g.][]{2019PhRvL.123c5101L, 2019PhRvE.100a3205P, 2019PhRvE.100c3209L, 2019PhRvE.100c3210L}. The microturbulence is mostly excited by the current filamentation instability (often termed Weibel), which generates an intense magnetic field on skin depth scales $\lambda_p$, with $\lambda_p \equiv c\,\left( 4\pi n_p e^2/m_p\right)^{-1/2}\sim 10^7\,n_{p\,0}^{1/2}\,$cm ($n_{p\,0}=n_p/1\,{\rm cm}^{-3}$). The effective magnetisation carried by this turbulence, written $\epsilon_B$, generally saturates at values $\epsilon_B\sim 0.01$ within thousands of skin depths from the shock front. Far downstream from the shock, it is expected to decay with distance as a mild power law~\citep{2013MNRAS.428..845L} through collisionless damping. Consequently, the effective magnetisation in the radiation region, which we write $\sigma_{\rm rad}$, is expected to be smaller than $10^{-2}$, but at least as large as the initial $\sigma$ value.

The modelling of gamma-ray burst afterglows offers an interesting perspective on the magnitude of $\sigma_{\rm rad}$. In those sources, the external magnetisation is significantly lower ($\sigma\sim 10^{-9}$) than  expected here, yet $\sigma_{\rm rad}$ is inferred to be as large as  $\sim10^{-5}- 10^{-4}$~\citep{2013MNRAS.435.3009L,2014ApJ...785...29S}. That effective magnetisation is generally understood as a left-over of the shock-generated field, or as a the result of further buildup through additional instabilities, for example the Richtmyer–Meshkov instability at the shock front~\citep{2011ApJ...734...77I} or a Rayleigh-Taylor instability at the contact discontinuity~\citep{2014ApJ...791L...1D}. The value of $\sigma_{\rm rad}$ that we use in our models is comparable to that above; in that sense, gamma-ray burst modelling provides empirical support to our model.

At moderate magnetisation, $10^{-5}\lesssim\sigma\lesssim10^{-2}$, a different current-driven instability is expected to contribute to the generation of the magnetised microturbulence~\citep{2014EL....10655001L}. The consequences of this instability, which is driven by the perpendicular current carried by the accelerated particles around the mean background magnetic field, have not been examined to the extent needed. In a first approximation, we can consider that the turbulence will have the same general characteristics as that generated by the current filamentation instability, that is, a strength $\epsilon_B\sim0.01$ and a length scale $\sim \mathcal O(\lambda_p)$, and we do so in the following.

The properties of the microturbulence in the vicinity of the shock govern the acceleration process, hence the parameters that characterise the spectrum of accelerated particles. In particular, the spectral index is expected to take values $s\simeq 2-2.3$. For definiteness, we assume $s\simeq2.2$ in our application to extreme-TeV blazars, corresponding to mildly relativistic shocks.
The minimum Lorentz factor is $\gamma_{e,\rm min}\simeq \overline\gamma_e\simeq 600\gamma_{\rm sh}$ as discussed before and the maximal Lorentz factor $\gamma_{e,\rm max}$ is determined by the conjunction of a number of constraints: (1) the age limit, $t_{\rm acc}\lesssim d/\left(\Gamma_{\rm j} \beta_{\rm j}c\right)$, where $t_{\rm acc}$ represents the acceleration timescale to $\gamma_{e,\rm max}$ and $d$ the distance from the jet base, $\Gamma_{\rm j}$ (resp. $\beta_{\rm j}$) the jet Lorentz factor (resp. velocity); (2) the radiative loss limit, $t_{\rm acc}\lesssim t_{\rm syn}$, written here as a competition with the synchrotron loss timescale; (3) the scattering limit (for super-luminal shocks) $t_{\rm scatt}\lesssim r_{\rm L,0}$, where $r_{\rm L,0}$ represents the particle gyroradius in the coherent (background) magnetic field. We note that the limit associated with lateral escape through the jet boundary, {\it viz.} $t_{\rm acc}\lesssim r$, is comparable to the age limit, if $r\sim \Theta_{\rm j}d$ and the jet opening angle $\Theta_{\rm j}\sim 1/\Gamma_{\rm j}$.

The acceleration timescale is given by $t_{\rm acc}\simeq t_{\rm scatt}$ in relativistic shocks, where $t_{\rm scatt}$ denotes the scattering timescale, which departs from the Bohm regime ($t_{\rm scatt}\propto r_{\rm L}$) at weak magnetisation~\citep[][and references therein]{2020Galax...8...33V}. Consequently, the radiative limit can be written $\gamma_{e,\rm max}\sim 10^7\left(n_e/1\,{\rm cm}^{-3}\right)^{-1/6}$, which does not depend on the strength of the magnetic field in the acceleration region, contrary to the usual relation $\gamma_{e,\rm max}\propto B^{-1/2}$ obtained for Bohm scaling. 

The scattering constraint ensures that particles are able to scatter effectively in the microturbulence, hence to cross the shock repeatedly, before being advected away from the shock by their gyration around the regular magnetic field lines~\citep{2009MNRAS.393..587P,2010MNRAS.402..321L}. This constraint thus assumes the magnetic field to be super-luminal and it would not arise in sub-luminal configurations. In this sense, it can be regarded as conservative. For scattering in microturbulence, this constraint can be reexpressed as a function of magnetisation:
\begin{equation}
\gamma_{e,\rm max}\lesssim \frac{\gamma_{e,\rm min}}{\sqrt{\sigma}}
\label{equ:gammamax}
.\end{equation}
Its main effect is to reduce the dynamic range over which particles can be accelerated. Its dependence on $\sigma$ derives from the scalings 
$t_{\rm scatt}\propto (\gamma_e/ \gamma_{sh})^2$ and $r_{\rm L,0}\propto (\gamma_e/\gamma_{sh}) \,/ \,\langle B\rangle$. The maximum Lorentz factor arising from this limit is multiplied by
$m_p/m_e$ in an electron-proton plasma. (See \citet{2020Galax...8...33V} for
more details.)

In the models that we adjust to the blazar SEDs further below, we find that this third constraint provides the limiting factor, although the radiative constraint and the age constraints do not lag by much. To put in some numbers, assume $\epsilon_B=10^{-2}$, $\sigma=10^{-5}$, $\gamma_{\rm sh}=3$, $n_e=1\,$cm$^{-3}$. Then, the scattering constraint gives $\gamma_{e,\rm max}\sim 2\times 10^5$, while the radiative and age constraints both lead to $\gamma_{e,\rm max}\sim 10^7$. The severity of the scattering constraint results from the relatively "low" value of $\gamma_{e,\rm min}$, here $1800$; below, we discuss models in which $\gamma_{e,\rm min}$ is larger because of acceleration at multiple shocks. In such cases, the maximal Lorentz factor that results from that scattering constraint becomes large enough to explain radiation up to TeV energies. Again, we stress that this scattering limit is conservative as it formally applies to super-luminal shocks and not sub-luminal ones.

We also modelled the radiation from protons, although we find that it is always negligible for the model parameters that we derive. The spectrum characterising the protons can be similarly defined by a minimum Lorentz factor, $\gamma_{p,\rm min}\sim \gamma_{\rm sh}$, a maximal Lorentz factor determined by the conjunction of the age and scattering constraints above and the same spectral index as for the electrons.

\subsection{Shock origin and geometry}

\subsubsection{Shocks of normal incidence}
\label{sec:shockvel}
The interaction of a jet with an obstacle generally triggers a double shock configuration, including a forward and a reverse shock. Depending on the relative momentum fluxes carried by the jet and the obstacle, either or both of these shocks can be strong or weak \citep{1995ApJ...455L.143S}. Given that particles release their radiation in the downstream rest frame of the shock where they are accelerated, the velocity of that downstream frame sets the Doppler factor that modulates the radiation. For extreme-TeV blazars, the large inferred Doppler boosting indicates that this downstream frame moves at large relativistic velocities towards the observer in the source frame. For shocks of normal incidence, this implies in turn that the shock itself moves at large velocities in the source frame, because the downstream rest frame moves at sub-relativistic velocities relative the shock. 

Consequently, if the jet overtakes an obstacle, we need to assume that the jet ram pressure far exceeds that of the obstacle, so that the Doppler boosting of the emission is given by that of the jet itself. The shock Lorentz factor $\gamma_{\rm sh}$ corresponds here to the relative Lorentz factor between the shock front and the obstacle, and it is of the order of $\Gamma_{\rm j}$ if the obstacle moves at slow velocities in the source rest frame. This may give rise to shocks of large Lorentz factors. However, obstacles can penetrate a jet only if their ram pressure exceeds that of the jet \citep[e.g.][]{2010A&A...522A..97A,2010ApJ...724.1517B,2012A&A...539A..69B}, which would instead imply a weak forward shock.

Alternatively, if an `obstacle' overtakes the jet, as in the `blob-in-jet' scenario, the Doppler factor will be at least as large as that of the jet itself. If the blob ram pressure exceeds that of the jet, the Lorentz factor of the emission region is given by that of the blob, $\Gamma_{\rm b}$, while the shock Lorentz factor $\gamma_{\rm sh}$ is set as before by the relative Lorentz factor between the blob and the jet.
From $\gamma_{\rm sh} =\Gamma_{\rm j} \Gamma_{\rm b} (1 - \beta_{\rm j} \beta_{\rm b})$,  
$\Gamma_{\rm j} \gg 1$, $\Gamma_{\rm b} \gg 1$ and $\Gamma_{\rm b} \gg \Gamma_{\rm j}$, it follows
\begin{equation}
\gamma_{\rm sh}\sim \Gamma_{\rm b}/(2\Gamma_{\rm j}) .
\label{equ:normalshock}
\end{equation}
Mildly relativistic shocks can be attained for sufficiently large ratios of $\Gamma_{\rm b}$ to $\Gamma_{\rm j}$. We consider such a configuration in Sect.~\ref{sec:sceII}.

\subsubsection{Oblique shocks}
\label{sec:oblique}
In the source rest frame, the shock(s) may well be of oblique incidence, especially if they are recollimation shocks~\cite[e.g.][and references therein]{2013EPJWC..6102002P,2016ApJ...831..163M}. The general properties of these recollimation shocks have been explored in various studies \citep[e.g.][]{1993MNRAS.261..203D,1997MNRAS.288..833K,2009MNRAS.392.1205N,2009ApJ...699.1274B}. We summarise in Appendix~\ref{app:recollimation} the characteristics of interest to us, in particular the jump conditions, and we establish here the correspondence between these properties and the above constraints on particle acceleration. 

In brief, these constraints remain applicable to oblique shocks once they are expressed in the so-called shock normal rest frame, in which the (un-shocked) flow impinges on the shock at normal incidence~\citep{1990ApJ...353...66B}. The Lorentz factor $\gamma_{\rm sh}$, which corresponds to the Lorentz factor of the shock with respect to the upstream plasma in this shock normal frame, is written $\Gamma_{\rm j<\vert n}$ in Appendix~\ref{app:recollimation}. It can be related to the relative Lorentz factor of the (pre-shock) jet with the (oblique) shock surface, written $\Gamma_{\rm j<}$, through Eq.~(\ref{eq:AC3}): $\gamma_{\rm sh}= \Gamma_{\rm j<}/\Gamma_{\rm n\vert s}$, where $\Gamma_{\rm n\vert s}$ represents the Lorentz factor associated with the boost to the shock normal frame. The main result here can be phrased in terms of the angle $\left\vert \theta_{\rm j<}-\alpha\right\vert$ between the flow incidence and the shock surface ($\theta_{\rm j<}$  and $\alpha$, respectively, represent the angles of the flow and the shock surface with respect to the jet axis). Given that the jet opening angle is of the order of $1/\Gamma_{\rm j}$, one expects $\left\vert \theta_{\rm j<}-\alpha\right\vert \sim 1/\Gamma_{\rm j}$ in a simple re-confinement scenario \citep{2009MNRAS.392.1205N}, and we therefore write $\left\vert \theta_{\rm j<}-\alpha\right\vert \sim \kappa/\Gamma_{\rm j<}$. Then, we obtain $\gamma_{\rm sh}\simeq \sqrt{1+\kappa^2}$. The shock is thus mildly relativistic for $\kappa$ of the order of unity. The angle $\left\vert \theta_{\rm j<}-\alpha\right\vert$ may however show a different behaviour in jets where recollimation shocks arise from a difference in pressure between jet components and the ambient medium \citep[e.g.][]{Fichet2020}.

Similarly, we can write the Lorentz factor of the flow past the shock as $\Gamma_{\rm j>} \simeq \Gamma_{\rm n\vert s}\simeq \Gamma_{\rm j<}/\sqrt{1+\kappa^2}$. This Lorentz factor is that which controls the amount of Doppler boosting from the emission region, provided the recollimation shock is stationary in the source frame. We note that the flow is refracted by an angle of the order of $\left\vert \theta_{\rm j<}-\alpha\right \vert$ in the source frame as it crosses the shock, which may therefore affect the overall Doppler boosting from the post-shock region quite substantially relative to that in the pre-shock region. We do not consider this effect in the models that follow, for the sake of simplicity. 

In our models below, we seek to describe the radiation emitted by an element of plasma crossing such a shock, or a series of shocks. In this description, we can model the plasma element as a blob containing a fixed number of particles, in which particles get accelerated at some stage then radiate. Even though we are dealing with a standing shock, which provides a steady emission pattern, the corresponding Doppler boosting of the radiation from the blob, as seen by the distant observer, is  a factor $\delta^4$ \citep[c.f.][]{1997ApJ...484..108S}, with $\delta$ the bulk Doppler factor. For the sake of completeness, we provide more details on this issue in Appendix~\ref{app:emission}.

\subsubsection{Interaction with multiple shocks}
\label{sec:reacceleration}
As recollimation (or, more generally, standing) shocks may come in series of shock fronts in the jets of radio-galaxies, we must consider the possibility that the radiating blob undergoes multiple episodes of shock acceleration, which may substantially modify the spectrum of accelerated particles \citep{1985ApJ...289..698W,1990A&A...231..251A,1993A&A...278..315S, 1994PASAu..11..175P,2013A&A...556A..88M}.  Particles are heated and accelerated in those shocks, but they also suffer adiabatic cooling (and possibly, radiative cooling) in the rarefaction regions that separate them.

As a particle crosses a shock front, its energy in the local frame of rest of the plasma is increased on average by the relative Lorentz factor between pre-shock and post-shock flows. Average is understood here over the orientation of the particle momentum in the pre-shock flow. For an oblique shock, the relative Lorentz factor $\Gamma_{\rm rel}$ can be written in terms of normal shock frame quantities $\Gamma_{\rm rel}\simeq 0.7\Gamma_{\rm j<\vert n}$, where the prefactor holds for mildly relativistic shocks with $\Gamma_{\rm j<\vert n}\gtrsim 2-3$. In terms of source frame quantities,
$\Gamma_{\rm rel} \simeq 0.5\Gamma_{\rm j<}/\Gamma_{\rm j>}$ (in analogy with Eq.~\ref{equ:normalshock}).

Adiabatic cooling in the rarefaction regions implies, for one-dimensional dilatation along the flow axis, $\Delta\ln p = -\frac{1}{3}\Delta \ln u_{\rm j}$, with $u_{\rm j}$ the flow four-velocity. Consider for instance a velocity profile such that $\Gamma_{\rm j<}$ decreases abruptly to $\Gamma_{\rm j>}$ through a recollimation shock, then steadily re-increases to its initial value $\Gamma_{\rm j<}$ through rarefaction. Overall, the momentum of the particle increases by $g\sim \Gamma_{\rm rel}^{2/3}$ between a time defined as immediately before the crossing of the first shock and that immediately before the crossing of the second shock. The difference with the sub-relativistic regime studied in the above references should be noted : there the particle gains little energy through shock crossing, but cools efficiently in the rarefaction region, leading to an overall energy loss in between two consecutive shock crossings. 

Beyond this sequence of heating and cooling, which shifts the energy spectrum in momentum space while preserving its shape, the particles are also re-accelerated at the shock. This now modifies the spectral shape. This effect can be described, in a first approximation, through the effective propagator
\begin{equation}
    G\left(\gamma_>,\,\gamma_<\right)=\frac{s-1}{g\,\gamma_<}\left(\frac{\gamma_>}{g\,\gamma_<}\right)^{-s}\Theta\left(\gamma_>-g\,\gamma_<\right)\,,
    \label{eq:AD3}
\end{equation}
which provides the probability distribution of the outgoing particle Lorentz factor $\gamma_>$ consequent to crossing a second shock, assuming a Lorentz factor $\gamma_<$ immediately after crossing a first shock. Here $g$ represents the energy gain from one shock to the next, $g\simeq\Gamma_{\rm rel}^{2/3}$ as discussed above. The post-shock particle distribution can be obtained through the convolution of the pre-shock distribution with this propagator, which gives, after $n$ shock crossings  \citep{1985ApJ...289..698W,1990A&A...231..251A,1993A&A...278..315S}
\begin{equation}
    \frac{{\rm d}N_>^{(n)}}{{\rm d}\gamma_>}\,=\,\frac{(s-1)^{n+1}}{n!\,g^n\,\gamma_{\rm min}}\left(\frac{\gamma_>}{g^n\,\gamma_{\rm min}}\right)^{-s}\ln\left(\frac{\gamma_>}{g^n\,\gamma_{\rm min}} \right)^n\,.
    \label{eq:AD5}
\end{equation}
This has two important consequences: (1) the spectrum hardens because of reacceleration; (2) the effective injection Lorentz factor, which was $\gamma_{\rm min}$ at the first shock, has become $g^n \gamma_{\rm min}$ at the $n-$th shock. The hardening is stronger at momenta close to the effective injection momentum than at high energy (HE). In particular, the maximum of ${\rm d}N_>^{(n)}/{\rm d}\ln \gamma$, which provides a reasonable estimate for $\overline\gamma_e$ is found at $\gamma\,=\,g^n\, \gamma_{\rm min} \exp\left[n/(s-1)\right]$, which indicates a rapid rise of the effective $\overline \gamma_e$ with the number of shock crossings, all the more so if $g$ is larger than unity ({\it i.e.} for relativistic shocks).

To make connection with the model parameters, $\gamma_{e,\rm min}$ is now given by $g^n\,\gamma_{e,\rm min}$, with $\gamma_{e,\rm min}\simeq 600\gamma_{\rm sh}$ and the spectrum is given by Eq.~(\ref{eq:AD5}) above.

\subsection{One-zone model with $e-p$ co-acceleration in relativistic shocks}
\label{sec:model}

The standard one-zone SSC model has usually nine free parameters: the strength of the uniform magnetic field $B$, the size of the spherical emission region $R$, its Doppler factor $\delta$, the minimum, maximum and break Lorentz factors of the electron distribution $\gamma_{\rm e,min}$, $\gamma_{\rm e,max}$, $\gamma_{\rm e,br}$, the two indices of the electron distribution before and after the break $s_1$ and $s_2$, and its normalisation $K_e$ at $\gamma = 1$\footnote{All model parameters refer to a frame co-moving with the emission region, i.e.\, the plasma `blob'.}. Even with a well-sampled multi-wavelength dataset, this model remains highly degenerate.

\cite{1998ApJ...509..608T} have studied the constraints on these parameters for a set of six observables extracted from the SED: the peak frequencies and luminosities of the synchrotron and inverse Compton (IC) components, $\nu_{\rm syn}$, $\nu_{\rm IC}$, $\nu_{\rm syn} L_{\rm syn}(\nu_{{\rm syn}})$, $\nu_{\rm IC} L_{\rm IC}(\nu_{\rm IC})$, and the indices in the differential photon flux $F_{\nu}$ before and after the synchrotron peak, $\alpha_1$ and $\alpha_2$. A detailed method for an exploration of this parameter space for a given dataset, allowing for statistical uncertainties in the data, has been discussed by \cite{2013A&A...558A..47C}.

The constraints on the magnetisation, on $\gamma_{e,\rm min}$, $\gamma_{e,\rm max}$ and on the spectral index, which we derived above from the microphysics of relativistic shock acceleration, remove most of the inherent degeneracy of the standard model. We demonstrate this by considering a power-law electron distribution with an exponential cutoff of the form 
\begin{equation}
\frac{dN_{e}}{d\gamma} = K_{e} \gamma^{-s} \, e^{-\gamma/\gamma_{e,\rm max}} \,.
\end{equation}
This simple parameterisation turns out to be well suited to characterise the SEDs of extreme-TeV blazars.

For an initially chosen value of $\delta$, first the maximum electron Lorentz factor $\gamma_{e,\rm max}$ can be determined from the IC peak frequency $\nu_{\rm IC}$. In the Klein-Nishina regime, which is relevant for the datasets we want to investigate in the following section, this relation is given following \cite{1998ApJ...509..608T}:
\begin{equation}
\gamma_{e,\rm max} \simeq  \frac{1}{g(\alpha_1, \alpha_2)}\frac{1 + z}{\delta} \frac{h \nu_{\rm IC}}{m_e  c^2} \,,
\label{eq:gmax}
\end{equation}
with
\begin{equation}
    g(\alpha_1, \alpha_2) = \exp \left( \frac{1}{\alpha_1 - 1} + \frac{1}{2(\alpha_2 - \alpha_1)} \right) \,.
\label{eq:gIC}
\end{equation}
Setting $s\simeq 2.2$ then implies $\alpha_1\simeq0.6$ as $\alpha = (s - 1) / 2$ because the particles do not effectively cool through synchrotron radiation. The value of $\alpha_2$ is fixed to 1.75 for all sources under study, which provides a sufficiently good approximation for the shape of the synchrotron spectrum just after the peak, when assuming an electron distribution with exponential cutoff. 

With $\gamma_{\rm e, max}$ known, the magnetic field strength $B$ in the radiation region can be estimated from the value of $\nu_{\rm syn}$, giving  
\begin{equation}
B = 27\, {\rm mG}\,\frac{1+z}{\delta}\frac{\nu_{\rm syn}}{10^{17}\,{\rm Hz}}\left(\frac{\gamma_{e,\rm max}}{10^6}\right)^{-2}\,.
\label{eq:Bvalue}
\end{equation}

 Since the above formulas were derived for a broken-power-law distribution, a small adjustment is needed for a good representation with our electron distribution. Multiplying the observed values of $\nu_{\rm syn}$ and $\nu_{\rm IC}$ by a factor 2.0 has been found to adjust well for the gradual flux decrease due to the exponential cutoff.

Provided the maximum electron energy is set by the scattering constraint, following Eq.~(\ref{equ:gammamax}), and the minimum electron energy can be estimated, at least broadly, from the low-frequency part of the SED and the spectral slope in the GeV range, the magnetisation of the plasma flow can be estimated as $\sigma\simeq \left(\gamma_{e,\rm min}/\gamma_{e,\rm max}\right)^2$.

The size of the homogeneous emission region is derived from the peak luminosities, following again \cite{1998ApJ...509..608T} in the Klein-Nishina regime:
\begin{equation}
  R = \left\{ 2 \, \frac{f(\alpha_1,\alpha_2)}{\delta^4}\frac{\left[\nu_{\rm syn} L_{\nu_{\rm syn}}\right]^2}{\nu_{\rm IC} L_{\nu_{\rm IC}}B^2 c}\,\left(\frac{3\delta ~ m_e ~ c^2}{4 \gamma_{e,max} ~ h  \nu_{\rm syn}} \right)^{(1-\alpha_1)} \right\}^{1/2} \,,
\label{eq:Req}
\end{equation}
with 
\begin{equation}
    f(\alpha_1, \alpha_2) = \frac{1}{1-\alpha_1} + \frac{1}{\alpha_2-1}\,.
\end{equation}

We may then verify our assumption of a single power-law distribution without cooling break,  by comparing the frequency at which synchrotron cooling becomes relevant, namely\ 
\begin{equation}
\gamma_{\rm c} \simeq 7.7 \times 10^8 \left( \frac{B}{1\,{\rm mG}} \right)^{-2} \left(\frac{R}{3 \times 10^{16}{\rm cm}} \right)^{-1}
\end{equation}
{\it versus} $\gamma_{e,\rm max}$. Given the low magnetisation required for efficient shock acceleration, the assumption holds in our application. IC cooling is not significant either in the absence of strong Compton dominance.

Hence, the only missing parameter is the normalisation of the electron distribution $K_e$. Following our scenario outlined in the previous section, a population of protons is added to the model. In a standard one-zone lepto-hadronic model, this would lead to four additional free parameters for a power-law distribution \citep[e.g.][]{2013A&A...558A..47C,Zech2017}. In our scenario, these parameters are constrained by the microphysics, as discussed earlier. 

The normalisation $K_e$ can then be determined from the magnetisation. We first determine the normalisation of the proton spectrum $K_p$ from 
\begin{equation}
    u_p \equiv \int_{\gamma_{p,\rm min}}^{\gamma_{p,\rm max}}{\rm d}\gamma\,  K_p \gamma^{-s+1} \, e^{-\gamma / \gamma_{p,\rm max}} \, m_p c^2  \, = \, \frac{u_B}{\sigma_{\rm rad}}\,,
\label{eq:uedens}
\end{equation}
with $\sigma_{\rm rad}$ the value of the magnetisation in the radiation region, which we recall is expected to be equal or larger than the pre-shock magnetisation. For a pure electron-proton plasma with equal particle number densities $n_e=n_p$, we then find 
\begin{equation}
   K_e  = K_p 
    \frac{\int_{\gamma_{p,\rm min}}^{\gamma_{p,\rm max}}{\rm d}\gamma\, \gamma^{-s} \, e^{-\gamma / \gamma_{p,\rm max}}}{\int_{\gamma_{e,\rm min}}^{\gamma_{e,\rm max}}{\rm d}\gamma\, \gamma^{-s} \, e^{-\gamma / \gamma_{e,\rm max}}}\,.
    \label{equ:ne_np}
\end{equation}
The normalisation $K_{e}$ does not {\it a priori} fit the absolute level of the observed flux defined by $\nu_{\rm syn} L_{\nu_{\rm syn}}$ and $\nu_{\rm IC} L_{\nu_{\rm IC}}$. Thus, in an iterative procedure, the value of $\delta$ needs to be varied to search for a solution for a given SED. The minimum allowed value of $\delta$ can be determined from constraints on the source opacity and from observations on the variability timescale, while its maximum value is more loosely limited by energetics and by the statistics of beamed and un-beamed sources~\citep{1998ApJ...509..608T}. 

In Eq.~(\ref{eq:uedens}) we have assumed that the proton and electron spectra share a same spectral index. This is a rather common assumption, of no consequence here, because the hadronic contribution to the radiation will be found to be undetectable in the present model. For reference, we assume $\gamma_{p,\rm min}= \gamma_{\rm sh}$ and $\gamma_{p,\rm max}= \gamma_{p,\rm min}/ \sqrt{\sigma}$, in line with the scalings adopted for the electrons, up to preheating in the shock transition. When considering a scenario where particles are re-accelerated on consecutive shocks, an analogous procedure can be followed by replacing the power-law particle distributions with the distribution given by Eq.~(\ref{eq:AD5}).

To enable a comparison of the energy requirements of the co-acceleration model to that of standard leptonic models, the jet power is evaluated using the usual approximation (for a two-sided jet):
\begin{equation}
    L_{\rm j} \approx 2 \pi R^2 \beta c \Gamma^2 (u_B + u_e + u_p + u_\gamma) \,,
\label{equ:Ljet}
\end{equation}
where $u_\gamma$ represents the energy density contained in the SSC radiation field.

We assume a value for the bulk Lorentz factor $\Gamma = 0.5 \, \delta$, corresponding to a situation where the jet is closely aligned with the line of sight. The required luminosity $L_{\rm j}$ needs to be compared to the theoretically available power from the central engine, for which the Eddington luminosity of the central black hole can be used as an order-of-magnitude estimate.

\section{Application to extreme-TeV blazars}
\label{sec:application}

The co-acceleration scenario was applied to five of the six multi-wavelength SEDs from extreme-TeV blazars presented by~\citet{Costamante2018} for the sources 1ES\,0229+200, 1ES\,1101-232, 1ES\,0347-121, RGB\,J0710+591, 1ES\,1218+304, and 1ES\,0414+009. These datasets were chosen due to the dense and homogeneous wavelength coverage including 
simultaneous data from {\it Swift} in the UV and soft X-ray band, from {\it NuStar} in the hard X-ray band and {\it Fermi} in the HE $\gamma$-ray band, together with contemporaneous
data at very high energy (VHE) from H.E.S.S. and VERITAS, as well as archival data from WISE in the infrared. It should be noted that, although the VHE data are not strictly simultaneous, extreme-TeV blazars are known for their lack of variability in the VHE range. Some level of variability has so far only been observed for 1ES\,0229+200 at a timescale of $\sim 10^7$\,s and 1ES\,1218+304 at a timescale of $\sim 10^5$\,s. 

The set of SEDs shows a clear continuity of the spectral shape between the HE and VHE range, except for the source 1ES\,0414+009, which exhibits a highly unusual spectral upturn between the {\it Fermi}-LAT and H.E.S.S. spectra. Such a concave shape, if confirmed, may indicate the presence of an additional spectral component, beyond the one-zone description we are providing here. It is however plausible, as assumed by~\citet{Costamante2018}, that a temporal variation in the spectral shape in the HE band is responsible for this apparent upturn between the non-simultaneous datasets. Given these uncertainties, this source has thus been excluded from our study, reducing the sample to five sources. 

As a first step, the values of the model parameters were determined using the analytical equations given in Sect.~\ref{sec:model}. A high Doppler factor of $\delta = 50$ was chosen initially following the results of~\citet{Costamante2018} and adjusted to lower or higher values if a good representation of the data could not be achieved.
The estimated parameter values were used as a starting point for modelling with a numerical code that treats the radiative transfer and all relevant leptonic and hadronic emission processes \citep[][]{Cerruti2015, Zech2017}. 
The contribution from the host galaxy was reproduced with a template from~\citet{1998ApJ...509..103S},  where the normalisation was adjusted to fit the infrared and optical data points. We have verified, analytically and with the time-dependent code by~\citet{Dmytriiev2020}, that in the resulting solutions, radiative cooling of the emitting particle population can be safely neglected, given the low magnetisation that is inherent to the shock acceleration scenario.

The co-acceleration model was declined into three scenarios to find the physically most 
meaningful representations of the selected SEDs. In a first scenario, we study a generic version with a large value of $\gamma_{e,\rm min}$, mimicking electron preheating at a shock of large Lorentz factor, or in some unspecified primary dissipation region. In a second scenario, we study the case of a lower value of $\gamma_{e,\rm min}$, as discussed earlier in the context of preheating at mildly relativistic internal or recollimation shocks. Finally, in a third scenario, we investigate the possibility of reacceleration at multiple shocks, focussing on the case of a single reacceleration at a secondary shock, sufficient to provide a good representation for the datasets under study, as will be seen.

In all cases, synchrotron radiation from the protons peaked at radio frequencies and was several orders of magnitude lower than the emission from electrons. It can thus be completely neglected when studying the SEDs. Only the kinetic energy of the protons needs to be considered when discussing the energetics of the source.

\subsection{Scenario I: Generic shock model with large $\gamma_{e,\rm min}$}\label{sec:sceI}
 We thus consider here a model with a large value of $\gamma_{e,\rm min}$. In light of our discussion in Sect.~\ref{sec:coacceleration}, such a large value can either represent the consequence of electron preheating at a shock of large Lorentz factor or some  pre-energisation in some dissipation process located well upstream of the shock where electron acceleration takes place. In the latter two-zone model, protons do not play any role at the shock; it would therefore apply equally well for a pure or high-multiplicity electron-positron composition. 

In the former, a large value of the shock Lorentz factor ({\it e.g.} $\sim\delta$), means that the jet encounters an obstacle moving at small velocity in the source rest frame (see Sect.~\ref{sec:shockvel}). However, this requires the jet ram pressure to exceed that of the obstacle in the source frame, which is precisely opposite to the condition that allows the obstacle to penetrate the jet \citep[e.g.][]{2010A&A...522A..97A,2010ApJ...724.1517B,2012A&A...539A..69B}.
Consequently, none of the above possibilities currently receives clear justification. For this reason, we regard this first scenario as a case study. To fix the parameters, we assume that $\gamma_{\rm e, min}$ is set by the shock Lorentz factor, in accordance with Sect.~\ref{sec:microphysics}, and that this shock Lorentz factor is set equal to the Doppler factor $\delta$, then slightly adjusted by constraints from the slope of the {\it Fermi}-LAT spectrum where necessary. This turned out to be necessary for one source only, 1ES\,1218+304 (see below). 

This choice leads to very high values of $\gamma_{\rm e, min}$, of a few $10^4$, in agreement with previous modelling attempts that can be found in the literature, where $\gamma_{\rm e, min}$ is treated as a free parameter.  We further assumed that the magnetisation in the emission region is the same as in the region upstream of the shock (i.e.\ $\sigma_{\rm rad} = \sigma$). This choice of constraints
removes the usual degeneracy of the SSC model. Ideally, following the procedure outlined in Sect.~\ref{sec:model}, only the Doppler factor $\delta$ is varied to adjust the model to the SED.

\begin{figure}[!h]
   \centering
   \includegraphics[width=\hsize]{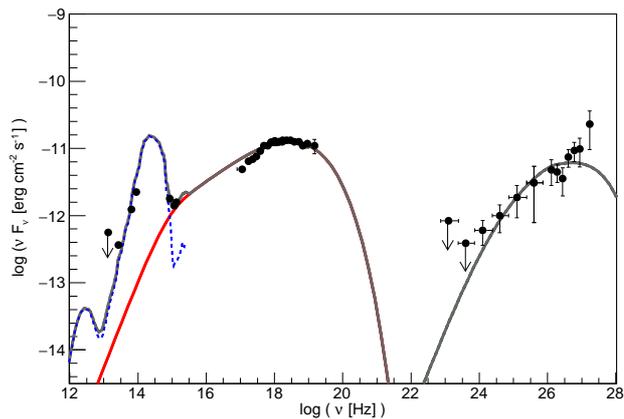}
      \caption{SSC model for the SED of 1ES\,0229+200 in scenario I. The data points at VHE are de-absorbed on the extragalactic background light (EBL) using the model by \citet{2008A&A...487..837F} in this figure and in all following figures showing SEDs.}
         \label{fig:sed0229}
   \end{figure}

\begin{figure}[!h]
   \centering
   \includegraphics[width=\hsize]{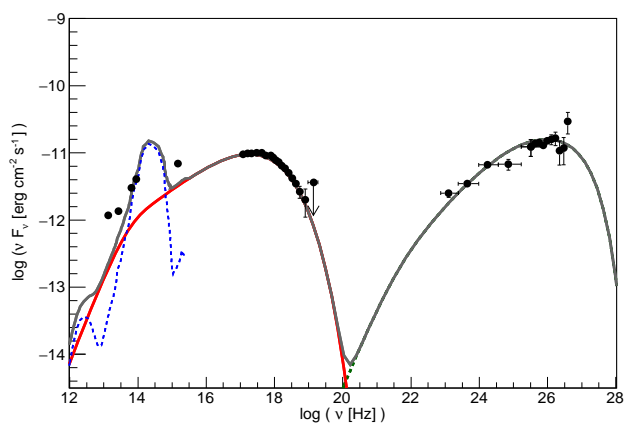}
      \caption{SSC model for the SED of 1ES\,1218+304 in scenario I. }
         \label{fig:sed1218}
   \end{figure}

Without any further adjustments, this scenario leads to a satisfactory representation of the SED of 1ES\,0229+200 for $\delta = 50$ (cf. Fig.~\ref{fig:sed0229}). Equally good solutions can be found for RGB\,J0710+591, 1ES\,0347-121, and 1ES\,1101-232 (cf. Fig.~\ref{fig:sed0710} to~\ref{fig:sed1101}) if, in addition to adjusting $\delta$ to values between $30$ and $50$, one also allows for an adjustment of the size of the emission region $R$, typically by a factor of $\sim1.5$. In the case of 1ES\,1218+304, the {\it Fermi}-LAT spectrum is flatter than
for the other sources, requiring a decrease in $\gamma_{\rm e,min}$ by an order of magnitude, in addition to adjustments of $\delta$ and $R$ (cf. Fig.~\ref{fig:sed1218}).

Allowing for these adjustments, satisfactory solutions can be found for all five SEDs. The full application of the constraints of the shock co-acceleration model leads to unique solutions in this generic scenario. As can be seen from Table~\ref{tab:par1}, the parameter values are very similar for the different sources: the Doppler factor of the emission region
is large (up to $\delta = 60$); the emission region has a size of a few $10^{16}$\,cm, a standard value in SSC models for high-frequency-peaked BL Lac objects (HBLs); the magnetic field strength, in the milligauss range, and the magnetisation are very small, as has been found before for these extreme sources. In our framework, a low magnetisation $\sigma\ll 10^{-2}$ is of course required to guarantee efficient acceleration.
   
In certain cases, for example the model of 1ES\,1218+304 shown in Fig.~\ref{fig:sed1218}, a mismatch between the model and the SED appears in the ultraviolet and infrared ranges.\ This may be explained with an additional synchrotron emission component from the extended jet, which is not represented in our models. 
   
\subsection{Scenario II: Co-acceleration on mildly relativistic shocks}\label{sec:sceII}  
In this second scenario, we consider the more physically grounded case of electron preheating and acceleration at internal (Sect.~\ref{sec:shockvel}) or recollimation shocks (Sect.~\ref{sec:oblique}).  In both cases, far smaller 
values of the shock Lorentz factor in the shock normal frame $\Gamma_{\rm j<\vert n}$ should be assumed than those applied in the generic scenario I. 

For acceleration on an internal shock caused by a plasma blob with Lorentz factor $\Gamma_{\rm b}$ traversing the jet, the resulting shock Lorentz factor
$\gamma_{\rm sh} \equiv \Gamma_{\rm j<\vert n}$ is given by Eq.~\ref{equ:normalshock}. Assuming a Doppler factor of $\delta$ for the observed emission from the blob, a mildly
relativistic shock with $\Gamma_{\rm j<\vert n} \simeq 3$ would result for a jet Lorentz factor $\Gamma_{\rm j} \simeq \delta / 12$, when assuming a closely aligned jet, corresponding to $\Gamma_{\rm b} \simeq \delta / 2$, such that $\Gamma_{\rm j}$ would be of the order of a few. Such a 
combination of a highly relativistic emission component responsible for HE emission and a mildly relativistic jet is frequently proposed to 
explain the discrepancy in observed bulk Lorentz factors in the radio and HE bands~\citep{2005A&A...432..401G, 1991ApJ...383L...7H, 1989MNRAS.237..411S}. In this case, shock Lorentz factors much higher than a few
can only be achieved for unrealistically high blob Lorentz factors or for non-relativistic jets. 

Considering the second case of acceleration on recollimation shocks, since the bulk Lorentz factor of the downstream, radiating plasma is $\Gamma_{\rm j>\vert s} \geq \delta / 2 $, the shock Lorentz factor must verify $\Gamma_{\rm j<\vert n} \simeq \Gamma_{\rm j<\vert s} / \Gamma_{\rm j>\vert s} \lesssim 2 \Gamma_{\rm j<\vert s} / \delta$ (cf. Appendix~\ref{app:recollimation}). If the bulk Lorentz factor of the plasma flow upstream of the recollimation shock $\Gamma_{\rm j<\vert s}$ is of a similar strength as the typical bulk Doppler factors derived from SED modelling, the shock Lorentz factor $\Gamma_{\rm j<\vert n}$ should be close to unity. The default value of $\Gamma_{\rm j<\vert n} = 3$ that we impose in this second scenario for both cases to better accommodate the datasets is already relatively high, but it should still be acceptable in this case.

With this assumption, the minimum electron Lorentz factor is fixed to $\gamma_{e,min} \simeq 600 \, \Gamma_{\rm j<\vert n}  \simeq 1800$.  The corresponding model fits imply different values of the physical parameters, following the procedure outlined in Sect.~\ref{sec:model}. Retaining bulk Doppler factors around $\delta \approx 50$, as for the solutions in scenario I, the decrease in $\gamma_{\rm e, min}$ for a fixed $\gamma_{\rm e, max}$ leads to a
quadratic decrease in the inferred upstream magnetisation $\sigma$ (cf. Eq.~(\ref{equ:gammamax})). Since the particle density is inversely proportional to the magnetisation downstream of the shock [cf. Eq.~(\ref{eq:uedens})], the condition $\sigma_{\rm rad} = \sigma$ needs to be abandoned to avoid a strong increase in the particle density, which would lead to a Compton dominance in the SED that is not observed in extreme blazars. 

An increase in the magnetisation in the region downstream of the shock by one or two orders of magnitude is not unexpected, due to the self-generation of magnetic turbulence at the shock. As discussed earlier, shock microphysics only imposes the condition $\sigma \leq \sigma_{\rm rad}\lesssim 0.01$, with $0.01$ representing the typical effective magnetisation at the shock, $\sigma_{\rm rad}$ the magnetisation in the radiation region and $\sigma$ the pre-shock magnetisation. We also stress that the values of $\sigma$ that are provided in Tab.~\ref{tab:par1} are only indicative, in the sense that they are derived from the respective values of $\gamma_{e,\rm min}$ and $\gamma_{e,\rm max}$ under the assumption that the maximal energy is set by the scattering constraint. As discussed earlier, this constraint applies formally to super-luminal shocks; it would be relaxed in sub-luminal shocks and, in such cases, the actual pre-shock magnetisation could be as large as $\sigma_{\rm rad}$. Consequently, the true physical magnetisation to be regarded is $\sigma_{\rm rad}$, of the order of $10^{-4}$ to $10^{-3}$ for the various sources in this model.

\begin{figure}[!h]
   \centering
   \includegraphics[width=\hsize]{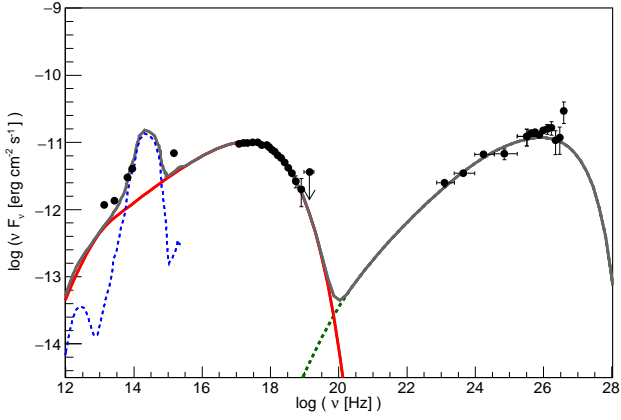}
      \caption{SSC model for the SED of 1ES\,1218+304 in scenario II.}
         \label{fig:sed1218_II}
   \end{figure}

As can be seen from Figures ~\ref{fig:sed1218_II}, ~\ref{fig:sed0710_II} and ~\ref{fig:sed0347_II}, satisfactory representations of the SEDs can be found within this  shock scenario for 1ES\,1218+304, RGB\,J0710+591 and 1ES\,0347-121. While for the first two, the representation of the SED is of comparable quality to the one in scenario I, for 1ES\,0347-121 the {\it Fermi}-LAT flux is over-predicted for the lowest flux point, but given the uncertainties in the {\it Fermi} points, this solution can still be regarded as acceptable. It should be noted here that, as far as the synchrotron and SSC peak flux levels are concerned,  the residual freedom in $\sigma_{\rm rad}$ introduces a degeneracy in the sense that similar solutions can be found by increasing $\delta$ and decreasing $\sigma_{\rm rad}$ or vice versa.

In the case of 1ES\,0229+200, the SSC component of the model is not sufficiently steep to pass through the $\gamma$-ray data points, as a result of the smaller value of $\gamma_{e,\rm min}$. 
The estimated SSC peak frequency, which is one of the input parameters for the modelling, needs to be increased significantly to be able to approach the VHE data points in 1ES\,0229+200. This leads to an increased radius of the emission region, of the order of a parsec, and a required jet power close to the Eddington limit. The magnetic field strength of this solution would be quite small, of the order of $10^{-5}$\,G.

The final source in the sample, 1ES\,1101-232, cannot be satisfactorily described either given the model constraints. In addition to a poor representation of the {\it Fermi}-LAT points, the infrared flux is overproduced by the model. If $\gamma_{\rm e, min}$ remains fixed at its default value, the only way around this problem is to increase $\delta$ to a value well above 100, in order to shift the low-energy break to sufficiently high frequencies.

Otherwise, satisfactory solutions for 1ES\,0229+200 and 1ES\,1101-232 can be obtained only if the shock Lorentz factor $\Gamma_{\rm j<\vert n}$ is increased to large values, thus recovering a scenario similar to I. These solutions are shown in Fig.~\ref{fig:sed1101_II} and~\ref{fig:sed0229_II} and the corresponding parameters are included in Table~\ref{tab:par1}. To reproduce the measured VHE flux in 1ES\,0229+200, a minimum value of $\Gamma_{\rm j<\vert n} \approx 20$ is necessary, while in the case of 1ES\,1101-232, $\Gamma_{\rm j<\vert n} \approx 10$ leads to a sufficiently steep rise of the SED at low energies to respect the constraints from the optical/IR data. 

Given these solutions, if acceleration were to take place at a mildly relativistic recollimation shock, this would lead to an initial bulk Lorentz factor of the jet of $\Gamma_{\rm j<\vert s} \approx 10^{3}$ for 1ES\,0229+200, while for 1ES\,1101-232, it would result in \ $\Gamma_{\rm j<\vert s} \approx 500 $. 
Such large jet Lorentz factors are problematic given that the usually assumed values for less extreme blazars, based on observations with very-long-baseline interferometry (VLBI) and on models of relativistic jets, are of the order of a few to a few tens~\citep{2009A&A...507L..33P,2016MNRAS.461..297S}.

For acceleration on internal shocks caused by blob-jet interaction, the required values of $\Gamma_{\rm j<\vert n}$ can be accommodated with jet Lorentz factors of a few, if the jet is
not too closely aligned with the line of sight (i.e.\ for $\delta \lesssim \Gamma_b$).

\subsection{Scenario III: Re-acceleration on a second shock}

Despite the difficulties  the recollimation shock scenario has in accounting for the SEDs with the hardest $\gamma$-ray spectra, it remains attractive as a natural explanation for continuous non-variable emission.
For this reason, we explore here the possibility that particles are accelerated at multiple standing shocks. As will be seen, such a possibility may help understand the peculiar SEDs of the most emblematic extreme blazars 1ES\,0229+200 and 1ES\,1101-232.

More generally, the need for continuous acceleration or multiple acceleration episodes along the jet is strongly supported by the non-thermal emission from radio galaxies at large (kpc) distances from the core, habitually observed in the radio, optical and X-ray bands, and recently even in VHE $\gamma$-rays~\citep{2020Natur.582..356H}. For simplicity, and to keep the model as economical as possible, we consider the simplest generalisation of scenario II, namely re-acceleration on a second shock of a particle distribution that had already been accelerated on an initial shock. As long as cooling remains negligible, re-acceleration leads to a  hardening particle spectrum with a low-energy turnover at increasingly higher energies above the initial value of $\gamma_{e,\rm min}$, as detailed in Sect.~\ref{sec:reacceleration}. 

The effect on the SED can be roughly estimated by approximating the electron distribution after $n$ re-accelerations with a hardening power law with increasing values of $\gamma_{\rm e,min}^{(n)} \simeq g^n \gamma_{\rm e,min}$. If the maximum energy is limited by the scattering constraint, it increases as well by $g^n$ (i.e.\ $\gamma_{\rm e,max}^{(n)} \simeq g^n \gamma_{\rm e,max}$), provided the magnetisation remains comparable from shock to shock. It cannot, of course, exceed the synchrotron or age limit. Fits of a power law to the distribution given by Eq.~(\ref{eq:AD5}), far from $\gamma_{\rm e,min}^{(n)}$ and assuming $g=2$ and a power law with index $s^{(0)} = 2.2$ for the initial acceleration, yield indices $s^{(1)} \simeq 2.1$ and $s^{(2)} \simeq 2.0$ for the first and second re-acceleration. In this approximation, the synchrotron and IC peak frequencies increase as $\nu_{\rm s}^{(n)} \propto g^{2n} \nu_{\rm s}$ and roughly $\nu_{\rm IC}^{(n)} \propto g^{n} \nu_{\rm IC} $, respectively.

The total emitted synchrotron power per unit volume and frequency at the frequency $\nu$ for the power-law electron distribution after $n$ re-accelerations in a turbulent magnetic field is then given following~\citet{Rybicki2004}:
\begin{equation}
\begin{aligned}
P_\mathrm{syn}^{(n)}(\nu) \propto \, & \, K_e^{(n)} \, \frac{1}{s^{(n)}+1} \, \Gamma \left(\frac{s^{(n)}}{4} + \frac{19}{12}  \right) \, \Gamma \left( \frac{s^{(n)}}{4} - \frac{1}{12}  \right) \\
 & \times \left( \frac{2 \pi \nu m_e c  }{3 e B}  \right)^{- (s^{(n)}-1)/2}
\end{aligned}
\label{eq:psyn}
,\end{equation}
with $K_e^{(n)}$ the normalisation of the power-law distribution. 

Keeping the electron number density and magnetic field strength constant, and estimating the 
peak frequency of the synchrotron component in the $\delta$ approximation as
$\nu_s^{(n)} \, {\rm [Hz]} \simeq 3.7 \times 10^6 \,B \, {\rm [G]} \, {\gamma_{\rm e, max}^{(n)}}^2$ ,
the increase in the emitted peak power for the first re-acceleration is roughly
\begin{equation}
\frac{\nu_s^{(1)} P_{\rm syn}^{(1)}(\nu_s^{(1)})}{\nu_s^{(0)} P_{\rm syn}^{(0)}(\nu_s^{(0)})}  ~ \simeq ~ 7.1 \, \left( \frac{ \gamma_{\rm e, max}}{10^{6}} \right)^{0.1} ~~.
\end{equation}
The numerical factor in the above equation depends weakly on the value of $\gamma_{\rm e, min}$, which we set again to $1.8\times10^3$ here.

Under the same assumptions, the Compton dominance of the emission (i.e.\ the ratio of the luminosities of the IC and synchrotron components) increases with re-acceleration. Following \citet{1998ApJ...509..608T}, it can be written as 
\begin{equation}
\begin{aligned}
\kappa^{(n)} \, \equiv \, \left( \frac{L_\mathrm{IC}}{L_\mathrm{syn}} \right)^{(n)} \,  \propto \, \, & f(\alpha_1^{(n)}, \alpha_2) \, \nu_{\rm s}^{(n)} \, P_\mathrm{syn}^{(n)}(\nu_{\rm s}^{(n)}) \, \\
&  \times \left( \frac{3 \, m_e c^2}{4 \times 3.7\times10^6 h}  \right)^{(3 - s^{(n)})/2} \,  \left( \frac{\delta}{B \, {\gamma_{\rm e,max}^{(n)}}^3} \right)^{(3 - s^{(n)})/2}   
\end{aligned}
.\end{equation}

The ratio of the Compton dominance between the first re-acceleration and the initial acceleration is then given by:
\begin{equation}
\frac{\kappa^{(1)}}{\kappa^{(0)}} ~\simeq ~ 2.6 \, \left( \frac{\delta}{50} \,
\frac{1 \, {\rm mG}}{B} \, \frac{10^{6}}{\gamma_{\rm e,max}}  \right)^{0.05} 
.\end{equation}

\begin{figure}[h]
   \centering
 \includegraphics[width=\hsize]{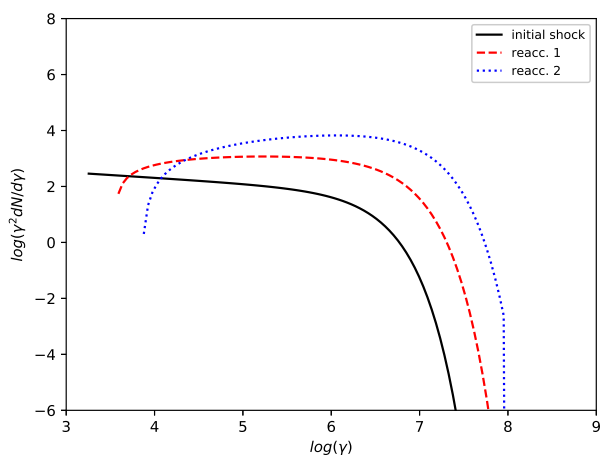}
      \caption{Particle spectra for the same electron and proton population after the initial shock acceleration and after a first and second re-acceleration on successive shocks.}
         \label{fig:particlespec}
   \end{figure}

\begin{figure}[h]
   \centering
   \includegraphics[width=\hsize]{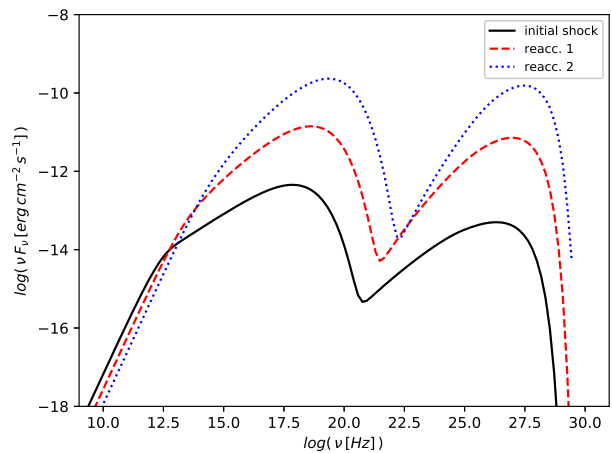}
      \caption{Observed SEDs due to SSC emission from the particle population shown in Fig.~\ref{fig:particlespec} after an initial shock acceleration and two re-accelerations. The SEDs correspond to a source at
      the redshift of 1ES\,0229+200 with $\delta = 50$.}
         \label{fig:SEDs}
   \end{figure}

The actual spectral distribution of re-accelerated particles deviates from a pure power law, leading to an even more significant effect on the
resulting SED. The evolution of a given electron population and its associated broadband emission from an initial shock acceleration to a first and second re-acceleration on similar shocks is shown in Figs.~\ref{fig:particlespec} 
and~\ref{fig:SEDs}. For this simulation, it was assumed that the number of particles remains constant during the evolution and that radiative cooling can be neglected. The source parameters $\delta$, $R$, and $B$ were adapted from the solution for 1ES\,0229+200 that will be presented below and were left constant for the consecutive accelerations. 
Under these admittedly simple assumptions, it can be seen that the resulting SED of the electron population is dominated by the last efficient shock acceleration.  

\begin{figure}[h]
   \centering
   \includegraphics[width=\hsize]{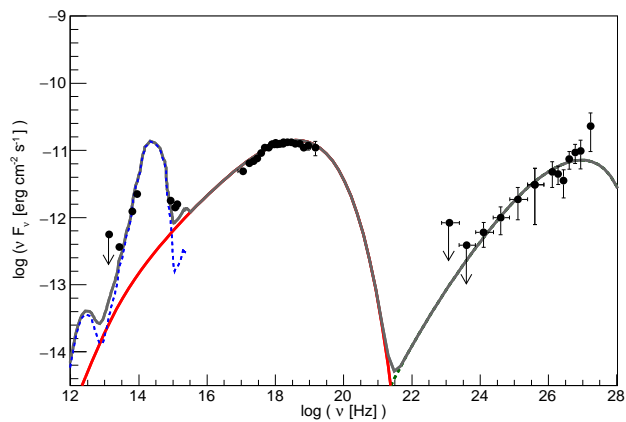}
      \caption{SSC model for the SED of 1ES\,0229+200 in scenario III.}
         \label{fig:sed0229_III}
   \end{figure}

\begin{figure}[h]
   \centering
   \includegraphics[width=\hsize]{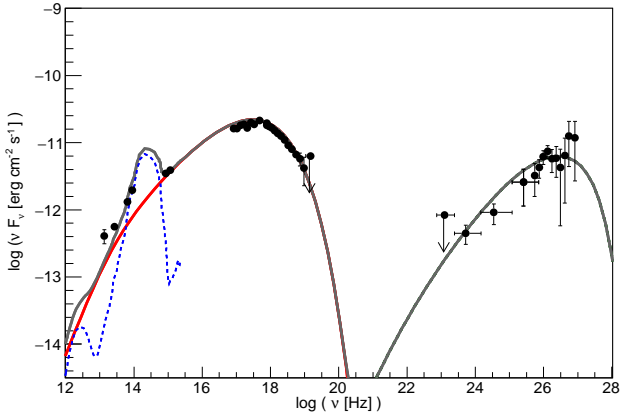}
      \caption{SSC model for the SED of 1ES\,1101-232 in scenario III.}
         \label{fig:sed1101_III}
   \end{figure}

Assuming that the emission from 1ES\,0229+200 and 1ES\,1101-232 is dominated by radiation of a particle population after a first re-acceleration, the SEDs of both sources can be well represented with parameters that are similar to scenario II; in particular a value of $\gamma_{\rm e,min}=1.8 \times 10^3$ is sufficient. To avoid an overestimate of the synchrotron peak frequency in the case of 1ES\,1101-232, due to the harder particle spectrum, the input value $\nu_{\rm syn}$ for the model was
reduced by a factor of 2. The results of this scenario are shown in Figs.~\ref{fig:sed0229_III} and~\ref{fig:sed1101_III}, with the parameter values given in Table~\ref{tab:par1}. This third
scenario also leads to a better representation of the SED of 1ES\,0347-121 (cf. Fig.~\ref{fig:sed0347_III}).

        \begin{table*}
   \centering
   \caption{Parameters used for the modelling of the extreme blazar datasets in scenarios I, II, and III.}
                \begin{tabular}{l c c c c c  }
                \hline
                \hline
                 \noalign{\smallskip}
                &1ES 0229+200 &1ES 0347-121  &RGB J0710+591 & 1ES 1101-232 & 1ES 1218+304\\
                 \noalign{\smallskip}
                \hline
                 \noalign{\smallskip}
         scenario I & & & & &\\
            \noalign{\smallskip}
                \hline
                 \noalign{\smallskip}
                $\delta $& 50 & 30 & 40 & 50 & 60 \\  
       $ R_{\rm src} \, [10^{16}\rm  cm]$ & 1.8 & 4.7 & 2.0 & 3.0 & 1.1 \\ 
                $B \, [\rm mG]$ & 4.4 & 3.7 & 7.9 & 8.9 & 8.4 \\  
                $\sigma_{\rm rad} \, (\, \equiv \sigma \,)$ & $1.0 \times 10^{-4}$ & $1.4 \times 10^{-4}$ & $5.0 \times 10^{-4}$ & $1.6 \times 10^{-3}$ & $9.8 \times 10^{-5}$ \\
                 \noalign{\smallskip}
                \hline
                 \noalign{\smallskip}
                $\gamma_{e,\rm min}$& $3.0 \times 10^4$ & $1.8 \times 10^4$ & $2.4 \times 10^{4}$ & $3.0 \times 10^4$ & $5.0 \times 10^3$ \\
                $\gamma_{e,\rm max}$ & $2.9 \times 10^{6}$ & $1.5 \times 10^6$ & $1.1 \times 10^6$ & $7.6 \times 10^{5}$ & $5.0 \times 10^5$ \\
            $n_e\ \textrm{[cm}^{\textrm{-3}}\textrm{]} $ & $2.9 \times 10^{-2}$ & $2.7 \times 10^{-2}$ & $2.8 \times 10^{-2}$ & $1.0 \times 10^{-2}$ & 0.69 \\
                 \noalign{\smallskip}
                \hline
                 \noalign{\smallskip}
                 scenario II & & & & & \\
                 \noalign{\smallskip}
                \hline
                 \noalign{\smallskip}
                $\delta $& 50 & 50  & 50  & 50  &  60 \\  
       $ R_{\rm src} \, [10^{16}\rm  cm]$ & 2.3 & 1.6 & 1.9 & 4.5  &  1.4 \\ 
                $B \, [\rm mG]$ & 4.4 & 6.2  & 9.8  & 8.9 & 8.4 \\  
                $\sigma_{\rm rad}$ & $1.9 \times 10^{-4} $ & $1.1 \times 10^{-4}$ &  $1.2 \times 10^{-3}$ & $4.0 \times 10^{-3}$ & $1.4 \times 10^{-4}$ \\
                $\sigma_{\rm rad} / \sigma$ & 11 & 29 & 270 & 64 & 11 \\
                 \noalign{\smallskip}
                \hline
                 \noalign{\smallskip}
                $\gamma_{e,\rm min}$& $1.2 \times 10^{4}$ & $1.8 \times 10^{3}$ & $1.8 \times 10^{3}$ & $6.0 \times 10^{3}$ & $1.8 \times 10^{3}$  \\
                $\gamma_{e,\rm max}$ & $2.9 \times 10^{6}$ & $9.1 \times 10^{5}$ & $8.6 \times 10^{5}$ & $7.6 \times 10^{5}$ & $5.0 \times 10^{5}$ \\
        $n_e\ \textrm{[cm}^{\textrm{-3}}\textrm{]} $ & $4.2 \times 10^{-2}$ & 0.83 & 0.24 & $2.1 \times 10^{-2}$ & 1.3 \\
                 \noalign{\smallskip}
                \hline
                 \noalign{\smallskip}           
                                 scenario III & & & & & \\
                 \noalign{\smallskip}
                \hline
                 \noalign{\smallskip}
                 $\delta $& 50 & 50 & - & 50 & -  \\ 
      $ R_{\rm src} \, [10^{16} cm]$ & 1.3 & 0.9 & - & 7.0 & - \\ 
                $B \, [\rm mG]$ & 4.4 & 6.2 & - & 4.5 & - \\  
                $\sigma_{\rm rad}$ & $1.0 \times 10^{-4}$  & $9.0 \times 10^{-5}$ & - & $4.0 \times 10^{-3}$ & - \\
                $\sigma_{\rm rad} / \sigma$ & 270 & 23 & - & 720 & - \\
                 \noalign{\smallskip}
                \hline
                 \noalign{\smallskip}
                $\gamma_{e,\rm min}$& $1.8 \times 10^{3}$ & $1.8 \times 10^{3}$ & - & $1.8 \times 10^{3}$ & - \\
                $\gamma_{e,\rm max}$ & $2.9 \times 10^{6}$ & $9.1 \times 10^{5}$ & - & $7.6 \times 10^{5}$ & - \\
        $n_e\ \textrm{[cm}^{\textrm{-3}}\textrm{]} $ & 0.13 & 0.29 & - & $8.2 \times 10^{-3}$  & \\
                 \noalign{\smallskip}
                \hline
                \hline          
                 \noalign{\bigskip}     
                \end{tabular}
         \label{tab:par1}
         \newline
         Note: The observables used as input for the modelling and parameters of the proton population can be found in Table~\ref{tab:par2}.
                \end{table*}

\section{Discussion}
\label{sec:discussion}

\subsection{Implications for the nature and location of the emission region}

The emission region described by the simple one-zone model may represent either a plasma blob moving with relativistic speed through the jet, into
which accelerated particles are injected from a leading bow shock \citep[similar to][]{1998A&A...333..452K}, or the region in the relativistic plasma flow downstream from a standing shock.

In both cases, one needs to explain why the emission is restricted mostly to a compact region, resembling the moving or standing VLBI radio knots observed in radio galaxies and blazars. Since in our scenarios synchrotron cooling is very slow, adiabatic cooling or particle escape into regions with even lower magnetic fields may be an explanation. As was discussed above, magnetisation is in fact expected to be amplified in the immediate neighbourhood of the shock. When particles escape from
the emission region into the downstream jet, their emission may be shifted to lower fluxes and frequencies. This low-energy emission from
the extended jet may be responsible for the radio flux, which we do not include in the SEDs shown here (cf. archival data in~\citet{Costamante2018}), and possibly contribute at infrared to ultraviolet frequencies, which may help to better reproduce the SED of
1ES\,1218+304 in this range (cf. Fig.~\ref{fig:sed1218_II}). Modelling of this extended 
emission component is beyond the scope of the present study.\\

 The blob-in-jet scenario can accommodate reasonably well a baryon loaded medium with low magnetisation if the blob interacts with the jet at large distances from the jet base. 
However, the blob emission timescale is constrained both by the expansion of the blob and its slowdown, due to its interaction with the jet. We assume that the expansion of the blob takes place at a lateral velocity $v_\perp$ in the source frame, with $v_\perp / c \sim \mathcal O(\Theta_{\rm j})$ if the expansion follows that of the jet~\cite[e.g.][]{2011AdSpR..48..998S}. Then
the available dynamical time in the blob frame  is  $t_{\rm dyn}'\sim R/(v_\perp\Gamma_{\rm b})$, corresponding to an observer timescale of $\Delta t_{\rm obs}=t_{\rm dyn}'/\delta \sim R/\left(v_\perp\Gamma_{\rm b}\delta \right)$. If $v_\perp/c\sim\Theta_{\rm j}$, the rather short dynamical timescale would imply substantial variations in the light curve on timescales shorter than a year. 
If the jet opening angle is smaller than the usual rough assumption (i.e.\ in the case of $\Theta_{\rm j} \ll 1/ \Gamma$ as discussed below), one could get a result closer to the observed slow or seemingly absent variability. In a transverse structured jet, the observed opening angle of the outer jet may also not be directly constraining the size of a blob that is propagating in an inner `spine'.

The blob slows down in the jet frame once it has swept up a mass $M_{\rm j} \sim E_{\rm b}/(\Gamma_{\rm b\vert j}^2c^2)$ , where $E_{\rm b}$ represents the blob energy (in the jet frame), and $\Gamma_{\rm b\vert j}$ the relative blob-jet Lorentz factor. On a timescale $t_{\rm dec}'$ (blob frame), the swept-up mass is $M_{\rm j}\sim \pi R^2 n_{\rm j} m_p \Gamma_{\rm b\vert j} c t_{\rm dec}'$, hence the deceleration time can be written $t_{\rm dec}'\sim \left[u_{\rm b}/\left(\Gamma_{\rm b\vert j}n_{\rm j} m_p c^2\right)\right]R/c$, with $u_{\rm b}$ the blob proper energy density. Consequently, $t_{\rm dec}'$ is generally larger than $t_{\rm dyn}'$, because the blob ram pressure $\Gamma_{\rm b\vert j}^2u_{\rm b}$ is assumed larger than $n_{\rm j}m_p c^2$, and $\Gamma_{\rm b\vert j}$ is of the order of a few at most.\\

The alternative scenario of acceleration on standing shocks, arising for example from recollimation, may account more naturally for a continuous emission.
In the standard picture of an initially Poynting-flux dominated jet, a first recollimation shock is assumed to occur at the transition between the magnetically dominated parabolic base of the jet and a kinetically dominated conical jet at the pc-scale~\cite[e.g.][]{2018MNRAS.473.4107P}. Oblique shocks due to collimation of the relativistic jet by the pressure of the surrounding medium have been considered by, for example, \citet{2009ApJ...699.1274B} to account for
the difference between large Doppler factors needed to explain TeV emission and the much smaller values inferred from unification models and 
super-luminal motion of radio knots. In this framework, acceleration and emission up to VHEs can occur up to several 10\,pc from the
base of the jet, from very compact emission regions with extensions down to $\sim 10^{-3}$ times the distance scale, in agreement thus with the size scale of the emission regions in our models. 

Direct observational evidence for the existence of such recollimation shocks has so far only been found from VLBI data in a few radio-loud AGN~\citep[e.g.][and references therein]{2018ApJ...860..141H}, and most prominently
in the most nearby radio galaxy M\,87 at the `HST-1' radio knot~\citep{2012ApJ...745L..28A,2013ApJ...775..118N}, which appears at a de-projected distance of approximately 300\,pc from the core, with a radius of the order of 1\,pc \citep{2012ApJ...745L..28A}. 

The typical extensions of the emission regions in our models are of the order of 0.01\,pc. To explain the HE emission from extreme-TeV blazars, much larger regions can be ruled out due to the energy requirements, as discussed in the case of 1ES\,0229+200 for our scenario II.
Besides the above-mentioned predictions from~\citet{2009ApJ...699.1274B}, an order-of-magnitude estimate of the location of the emission regions in our models can be made by comparing their radii with the assumed jet opening angle $\Theta_{\rm j}$, since $\Theta_{\rm j} \sim R/d$, with $d$ the distance of the emission region from the base of the jet. Typical values for $\Theta_{\rm j}$, based on VLBI data from the MOJAVE programme for $\gamma$-loud blazars are found to follow roughly $\Theta_{\rm j} \simeq 0.3  /\Gamma_{j|<s}$~\citep{2009A&A...507L..33P}, while~\citet{2015MNRAS.451..927Z} find
$\Theta_{\rm j} \simeq 0.1 /\Gamma_{\rm j}$ in a different sample. The latter note however that the opening angle is related to the magnetisation as $\Theta_{\rm j} \simeq s \sigma^{1/2}/\Gamma_{j|<s}$ with $s \lesssim 1$, which would suggest a relation closer to $\Theta_{\rm j} \simeq 0.01   /\Gamma_{j|<s}$ for the typical values of $\sigma$ found in our scenarios \footnote{
We note that their definition of $\sigma$ is slightly different from ours, which should however lead to a result of the same order of magnitude.}. These estimates lead to a range of distances from the core of the order of one to several tens of parsecs. 

Re-acceleration on successive standing shocks requires the distance between two shocks to be small compared to the cooling time of particles inside the plasma flow. 
Following Eq.~(\ref{eq:cooling_length}), the most energetic electrons in our models will not significantly cool at distance scales up to about $0.1 \, \Gamma_{\rm j|<s}$\,pc ({\it i.e.} a few parsecs).
This estimate is based on the magnetic field strength in the emission region, which is expected to be amplified compared to regions far from the recollimation shocks; the actual cooling distance may thus be far greater. Formation of internal shocks due to reflections of the converging recollimation shock, proposed by~\citet{2009ApJ...699.1274B}, may constitute alternative sites of re-acceleration.

\subsection{Implications for jet power, jet content, and magnetisation}

Contrary to other lepto-hadronic models in the literature, which try to explain the observed SEDs as a combination of radiative emission from electrons and protons, emission from protons is negligible in the
scenarios proposed here, due to the low magnetic field strengths and small maximum proton Lorentz factors $\gamma_{\rm p,max}$. Typical lepto-hadronic models achieve a significant radiative contribution from protons by assuming an acceleration process that is much more efficient for protons than for electrons, leading to $u_p \gg u_e$ and $\gamma_{\rm p,max} \gg \gamma_{\rm e,max}$, combined with high magnetic field strengths, of the order of 1\,G to a few 100\,G in the case of proton-synchrotron dominated emission. In such  models, the jet power is strongly dominated 
by the contribution from either the magnetic field or the relativistic protons, depending on the dominant hadronic emission mechanism, and is generally close to or even above the Eddington limit~\citep[e.g.][...]{2001APh....15..121M, 2012MNRAS.426..462P, Cerruti2015}. 

In our solutions, the magnetic field strength is several orders of magnitude smaller, the number densities of accelerated electrons and protons are equal ($n_e = n_p$), and the jet power is typically a factor $10^2$ to $10^3$ below the Eddington limit, closer to the results for BL Lac objects seen in the leptonic framework \citep[e.g.][]{2011MNRAS.414.2674G}. By design, emission emerges from a region of the jet with a low magnetisation $\sigma \ll 10^{-2}$.  

Studies of luminous, FSRQ-type blazars show that, at least for those objects, the assumption of a pure electron-proton jet is problematic, since required jet powers are very large compared to the available accretion power and compared to the power that is injected into radio lobes. The addition of an electron-positron pair plasma can lower the required jet power~\citep[e.g.][]{2016Galax...4...12S}. If this is the case, $n_e > n_p$, which would change our constraint on the normalisation of the electron spectrum $K_e$ (Eq.~\ref{equ:ne_np}). The value of $\gamma_{\rm e,min}$ is expected to decrease by the factor of electron multiplicity. A large multiplicity factor would thus exclude a good representation for the scenarios without substantial re-acceleration.  

The magnitude of the magnetisation $\sigma$ of blazar jets is still an open question. When assuming the Blandford-Znajek process for jet launching, this implies an initially strongly Poynting-flux dominated jet, in which magnetic energy is converted to kinetic energy over subparsec to parsec scales. During this conversion, the bulk Lorentz factor increases and the magnetisation drops to unity and below. In the standard picture, this conversion may proceed through differential collimation down to equipartition (i.e.\  $\sigma \simeq 1$), where it becomes inefficient. 

Further conversion may be driven by non-standard mechanisms, such as reconnection on magnetohydrodynamic (MHD) instabilities~\citep{2011MmSAI..82...95K}. In the latter case, a magnetisation $\sigma \ll 1$ seems reachable even at small distances from the base of the jet, where most of the radiation is expected to be produced.

In this case, \citet{2015MNRAS.451..927Z} show that blazar jets can be powered by the Blandford-Znajek mechanism, originate from magnetically arrested disks and can still have magnetisations $\sigma \ll 1$ at the radio core. They also find that in this framework jets are relatively heavy, with their mass dominated by baryons.

Such reconnection processes could lead to particle acceleration. However, if the magnetic field strength
in the transition region around $\sigma \lesssim 1$ does not exceed values of the order of 1\,G, the emission from those particles would contribute mainly to the extended radio and optical emission of the jet, not to the high-frequency part, because the average electron Lorentz factor decreases with decreasing $\sigma$ and the produced spectra become quite soft for $\sigma \lesssim 1$ (see Appendix~\ref{app:reconnection}). Such an additional emission component might actually help explain the discrepancy between our model and certain SEDs at low energies.

 A very rough comparison between the synchrotron peak emission close to such a transition region and the one from the HE emission zone can be made by considering two spherical regions with, in one case, $B_{\rm trans} \lesssim 1$\,G, $\sigma_{\rm trans} \lesssim 1$ and a steep particle spectrum with $\gamma_{\rm e,min} \sim 10^2$ and, in the other case, average conditions from our results from scenario II. Estimating
the peak luminosities using Eq.~\ref{eq:psyn} leads to a ratio of roughly $L_{\rm syn,trans} / L_{\rm syn,he} \sim 50 \, \left(R_{\rm trans}/10^{16}\,cm \right)^3 \left( B_{\rm trans} / 1 \, {\rm G} \right)^4 
\left(1 / \sigma_{\rm trans} \right)$ , with $R_{\rm trans}$ the radius
of the transition region. For sufficiently small values of $B_{\rm trans}$ and $R_{\rm trans}$, this emission may be hidden below that of the host galaxy and HE region. 

The luminosity of the IC peak from such a transition region would be much weaker than the synchrotron luminosity due to the steep particle spectrum. 
A more thorough treatment of this question would require simulations with a multi-zone jet emission model including a realistic description of the evolving reconnection regions, which is well beyond the scope of the current work.

In alternative scenarios where jet launching is described through the Blandford-Payne mechanism, a lower initial magnetisation of the jet may 
arise more naturally \citep[e.g.][for simulations in the non-relativistic limit]{2009MNRAS.400..820T}.

Direct measurements of the magnetic field strength $B$ in relativistic jets of blazars and radio galaxies are difficult. They usually rely on measures of the core shift, that is to say,\ the frequency-dependent position along the jet where synchrotron self-absorption becomes negligible and emission from the VLBI radio core can first be observed. To determine the value of $B$, usually equipartition between magnetic and particle energy densities is assumed and values of the order of 100\,mG are found at a distance of 1\,pc from the base of the jet~\citep{2010ASPC..427..207O,2014MNRAS.437.3396K}, while \citet{2017MNRAS.469..813A} find values of the order of 1\,mG at the more distant radio cores. Uncertainties and fluctuations in these measurements are however large and deviation from the assumption of equipartition will lead to different values.

\subsection{Applicability to other blazar types}

The SEDs of less extreme HBLs are well represented with smaller values of $\gamma_{e,\rm min}$, typically of the order of 1 to 100. These more common objects present at least an order of magnitude lower synchrotron and SSC peak frequencies than the extreme-TeV blazars, which translates into lower values of $\gamma_{\rm e,\rm max}$ and $\delta$, while the magnetic field strength is usually found to be at least an order of magnitude higher. 

In the present framework, the continuous emission from such objects could in principle be accounted for with lower Lorentz factors in the shock normal
frame. A higher electron multiplicity would also lead to lower values of $\gamma_{\rm e,min}$. Lower values of $\delta$ may be due to smaller Lorentz factors in the post-shock region or larger viewing angles of these objects. To reproduce the wider synchrotron and SSC bumps seen in many HBLs, particle spectra require in general a slightly more complex description, usually including a spectral break that cannot
be ascribed to radiative cooling of a single electron population alone. These features could possibly be reproduced with a combination of emission from several standing shocks (e.g.\ a strong first shock) followed by a weaker second shock. In the blob-in-jet scenario, the assumption of a uniform particle density and magnetic field strength in a spherical blob may be too approximate for a realistic description of the particle 
spectra, hence requiring additional degrees of freedom in the form of
an {\it ad hoc} spectral break. The acceleration process may also be more complex than what is assumed in our current model.

Contrary to extreme TeV blazars, flux and spectral variability are ubiquitous in HBLs, at timescales that can be as short as a few minutes in the observer frame at TeV energies. 
Additional mechanisms, possibly the interaction of moving shocks with successive standing shocks \citep[e.g.][and references therein]{2019ApJ...877...26H,Fichet2020} need to be considered to account for the appearance of high flux states and flaring events in blazars and radio galaxies.
Alternatively, depending on the magnetisation in the emission region, 
scenarios based on particle acceleration through reconnection or turbulence may provide
a suitable description for the variable emission from such objects.

\section{Conclusions}

Ascribing the non-thermal multi-wavelength emission from extreme-TeV blazars 
to the acceleration of a proton-electron plasma on mildly relativistic shocks 
introduces additional constraints, arising from the microphysics in the shock region, on the standard one-zone SSC model that is generally applied to describe SEDs of BL Lac objects. A more physically motivated description of the population of radiating electrons can thus be achieved.
The co-acceleration of electrons and protons results in a transfer of energy to the electrons, thus shifting their spectral distribution to higher energies. This effect provides a natural explanation for the high values of the minimum electron Lorentz factors required to model the SEDs of extreme-TeV blazars.

A first generic model, allowing for a very high bulk Lorentz factor of the jet
plasma upstream of the shock, leads to unique solutions for the set of five extreme-TeV blazar SEDs under study. 
In a second, more realistic scenario, exploring the two cases of acceleration on recollimation shocks or on the shock wave caused by a plasma blob propagating inside the jet, a good representation is found for most of the SEDs. High minimum Lorentz factors of the electron distribution of $\gamma_{\rm e, min} \gtrsim  10^{3}$ arise as a natural consequence of the assumed microphysics.

For the SEDs of the two objects with the hardest $\gamma$-ray spectra, 1ES\,0229+200 and 1ES\,1101-232, good representations are found for acceleration on moving shocks, but also for the acceleration on recollimation shocks when assuming that the particle distribution that dominates the emission results from re-acceleration on a second shock. In the latter case, the spectral hardening and the increase in the minimum and maximum Lorentz factors arising from the re-acceleration of an initial power-law electron distribution on a second shock can satisfactorily explain the extreme values of the synchrotron and SSC peak frequencies and the hard $\gamma$-ray spectrum.

\begin{acknowledgements}
The authors wish to thank Luigi Costamante for providing the data points of the SEDs and Zakaria Meliani for useful exchanges.     
\end{acknowledgements}

%
%

\begin{appendix}

\section{Electron heating in reconnection layers or magnetised turbulence}
\label{app:reconnection}

To produce electron spectra with large $\overline\gamma_e$, reconnection or turbulent acceleration must proceed in the relativistic regime, because they need to dissipate an energy reservoir $\sim\overline\gamma_e/\gamma_0$ times larger than the initial electron energy content, as discussed earlier. In a pure pair plasma, this would mean an initial magnetisation $\sigma\sim\overline\gamma_e/\gamma_0\gg1$, clearly at odds with the apparent requirement of a sub-equipartition magnetic field in the radiation region. As mentioned earlier, one may contemplate the possibility that such values are reached at the core of the jet in a region of high magnetisation, provided the radiation produced in this region does not dominate. Such dissipation generally leads, however, to an approximate equipartition between particles and magnetic fields, leading to a magnetisation $\sigma\sim\mathcal O(1)$, which remains too large to account for the spectral properties of extreme blazars, whence the need for some additional mechanism to bring down this magnetisation to the desired sub-equipartition level by the time the electron power law is shaped.

In an electron-proton plasma, the initial energy density is carried by the protons, hence reconnection can proceed in the relativistic regime from the point of view of electrons ($\sigma_e>1$, but in the sub-relativistic regime from the point of view of ions ($\sigma_p\lesssim 1$). Here, $\sigma_e$ (respectively, $\sigma_p$) denotes the electron (respectively, ion) magnetisation, that is, the ratio of the energy density in the magnetic field to the electron (ion) initial energy density.  The total magnetisation $\sigma$ is related to these two quantities via $\sigma^{-1}=\sigma_e^{-1}+\sigma_p^{-1}$. The simulations of \cite{2019ApJ...880...37P}, although conducted in the regime $\sigma\gtrsim 1$, suggest $\overline\gamma_e\sim 1+0.03\sigma^{3/2}m_p/m_e$ for an initially cold electron population, hence values $\sigma\gtrsim 10$ appear needed to reach $\overline\gamma_e\sim 10^3$. This result remains essentially unchanged if the electrons are initially hot ($T_e\gg m_e$). At lower magnetisations, the electron spectra take a more complex form, composed by part of their initial Maxwellian distribution at low Lorentz factors and a non-thermal population emerging out of a secondary Maxwellian at large Lorentz factor $\gamma_e^{\rm hi}$. From the results of \cite{2018ApJ...862...80B} and  \cite{2018MNRAS.473.4840W}, we infer $\gamma_e^{\rm hi}\sim \left(0.1-0.3\right)\sigma m_p/m_e$. Since $\overline\gamma_e\lesssim\gamma_e^{\rm hi}$ for such electron spectra, we recover the above result: in other words, $\sigma\gg 1$ is required to obtain $\overline\gamma_e\sim 10^3$. 

\cite{2018MNRAS.473.4840W} measure the fraction of dissipated energy $q_e=e_e/(e_e+e_p)$, where $e_e$ and $e_p$ are understood as kinetic energies. They obtain  
$q_e\sim 0.25 + 0.25\sigma^{1/2}/\left(10 + \sigma\right)^{1/2}$ in the regime $\sigma>0.01$, which, when combined with $\overline\gamma_p\simeq 1+0.5\sigma$ at $\sigma\lesssim 1$ \citep{2018ApJ...862...80B}, gives $\overline\gamma_e\simeq 0.1 \sigma m_p/m_e$ and thereby confirms  the previous estimate. 

Finally, we note that the slope of the accelerated spectrum depends rather sensitively on the magnetisation, for example $s\simeq4$ for $\sigma \lesssim 1$ and becoming as hard as $s\sim 2$ for $\sigma \gg 1$. All this clearly indicates that reconnection does not fulfil the requirements for a successful model for extreme blazars under our present assumptions.

The partition of energy between electrons and ions in MHD turbulence is a rather convoluted problem, which depends on a number of parameters, not only the initial magnetisation and species temperatures, but also the nature (compressive or Alfv\'enic) of the turbulence itself~\citep[e.g.][]{2020PhRvX..10d1050K}. As of today, there exist only a few kinetic simulations of ion-electron turbulence in the regime of interest and we rely on the results of \cite{2019PhRvL.122e5101Z}. These authors have performed a scan of simulations which span a broad range of magnetisations, from $\sigma_p\sim 10^{-3}$ (hence, $\sigma_e\sim 1$) up to the fully relativistic regime $\sigma_p\gtrsim 1$ ($\sigma_e\sim 10^3$). They observe that ions tend to be preferentially heated over electrons and that the amount of electron heating is found to scale with the initial temperature. For an initial electron temperature $T_e \simeq m_e$, the electrons acquire about 10\% of the ion energy gain. If $\gamma_p-1\sim\sigma$, as expected for near-complete transfer of the magnetic energy into particles, this means $\overline\gamma_e\simeq 0.1\sigma m_p/m_e$, a result comparable to what is observed in reconnection layers. Hence, $\overline\gamma_e\gtrsim 10^3$ can be achieved for large magnetisations only.

\section{Jump conditions at a recollimation shock}
\label{app:recollimation}
Here we concentrate on the shock crossing conditions and rewrite them in a simple way, in order to derive the post-shock flow velocity and to keep the discussion self-contained. We do so, in particular, by de-boosting to the so-called `shock normal frame' $\mathcal R_n$ \citep{1990ApJ...353...66B} in which the flow moves along the shock normal. This means de-boosting by the component of the velocity which is transverse to this shock normal (equivalently, parallel to the shock surface).

We assume that the surface of the recollimation shock makes an angle $\alpha$ with respect to the jet axis. The flow direction, at angle $\theta_{\rm j<}$ with respect to this jet axis, thus forms an angle $\pi/2 - \left(\theta_{\rm j<}-\alpha\right)$ with respect to the shock normal. The symbol $_{<}$ is used to distinguish pre-shock from post-shock quantities (indexed with $_>$). If the shock front is stationary in the source rest frame $\mathcal R_{\rm s}$, then the shock normal frame $\mathcal R_{\rm n}$ moves at velocity $\boldsymbol{\beta_{\rm n\vert s}}$ with respect to $\mathcal R_{\rm s}$, where
\begin{equation}
    \boldsymbol{\beta_{\rm n\vert s}}=\boldsymbol{\beta_{\rm j<}}\,-\,
    \left(\boldsymbol{\beta_{\rm j<}}\cdot \boldsymbol{n_{\rm\vert s}}\right)\boldsymbol{n_{\vert\rm s}}\,.
    \label{eq:AC1}
\end{equation}
All throughout $\boldsymbol{n_{\vert\rm s}}$ represents the (unit) direction of the normal to the shock front in the source rest frame. This gives $\beta_{\rm n\vert s}=\beta_{\rm j<}\cos\left(\theta_{\rm j<}-\alpha\right)$, hence $\Gamma_{\rm n\vert s}=\left[1-\beta_{\rm j<}^2\cos^2\left(\theta_{\rm j<}-\alpha\right)\right]^{-1/2}$. 

To derive the effective velocity of the jet in this shock normal frame, we note that the component of the jet four-velocity along the shock normal is preserved by the de-boost, hence $u_{\rm j<\vert n} = \Gamma_{\rm j<}\beta_{\rm j<}\sin\left(\theta_{\rm j<}-\alpha\right)$. The relative Lorentz factor between the recollimation shock and the incoming plasma in the shock normal frame, which matters for acceleration purposes, can thus be written 
\begin{equation}
\Gamma_{\rm j<\vert n}=\sqrt{1+\Gamma_{\rm j<}^2\beta_{\rm j<}^2\sin^2\left(\theta_{\rm j<}-\alpha\right)}\,=\,\frac{\Gamma_{\rm j<}}{\Gamma_{\rm n\vert s}}\,.
\label{eq:AC3}
\end{equation}

If $\Gamma_{\rm j<}\left\vert\theta_{\rm j<}-\alpha\right\vert\ll 1$, the recollimation shock is sub-relativistic in the $\mathcal R_n$ frame, otherwise it is relativistic.
The post-shock downstream velocity can be approximated (in $\mathcal R_{\rm n}$) as $\beta_{\rm j>\vert n}\simeq \beta_{\rm j<n}/r$ in the weakly magnetised limit, with $r$ the compression ratio. Hereafter we assume a mildly relativistic shock with $r\simeq 3$, implying $\beta_{\rm j>\vert n}\simeq 0.3$, hence $\Gamma_{\rm j>\vert n}\sim 1$. For reference, the exact shock jump conditions for a strong hydrodynamical shock give $\beta_{\rm j>\vert n}= 0.29$ if $\Gamma_{\rm j<\vert n}=3$ , $\beta_{\rm j>\vert n}=0.26$ if $\Gamma_{\rm j<\vert n}=2$.

Consequently, boosting back to the source frame gives a post-shock plasma four-velocity
\begin{equation}
    \boldsymbol{u_{\rm j>\vert s}}\,=\,\Gamma_{\rm j>\vert n}\beta_{\rm j>\vert n}\boldsymbol{n_{\vert\rm s}}\,+\,\Gamma_{\rm j>\vert n}\Gamma_{\rm n\vert s}\boldsymbol{\beta_{\rm n\vert s}}\,,
    \label{eq:AC4}
\end{equation}
with corresponding Lorentz factor 
\begin{equation}
\Gamma_{\rm j>\vert s}=\Gamma_{\rm j>\vert n}\Gamma_{\rm n\vert s}\,\simeq\,\frac{\Gamma_{\rm j<}}{\sqrt{1+\beta_{\rm j<}^2\Gamma_{\rm j<}\sin^2\left(\theta_{\rm j<}-\alpha\right)}}\,.
\label{eq:AC5}
\end{equation}

If $\Gamma_{\rm j<}\left\vert\theta_{\rm j<}-\alpha\right\vert\ll 1$, the shock is weak, as noted above, and the post-shock plasma flows nearly along the shock surface. In the opposite limit, the recollimation shock is ultra-relativistic in $\mathcal R_{\rm n}$ (see Eq.~(\ref{eq:AC3})), the shock is strong and the post-shock flows at sub- or mildly relativistic speeds in $\mathcal R_{\rm s}$ (see Eq.~(\ref{eq:AC5})). To describe the intermediate limit, write $\left\vert\theta_{\rm j<}-\alpha\right\vert = \kappa/\Gamma_{\rm j<}$ with $\kappa\sim\mathcal O(1)$. Then $\Gamma_{\rm j<\vert n}\simeq \sqrt{1+\kappa^2}$ and $\Gamma_{\rm j>\vert s}\simeq \Gamma_{\rm j<}/\sqrt{1+\kappa^2}$. 

We also note that Eqs.~(\ref{eq:AC3}) and (\ref{eq:AC5}) give $\Gamma_{\rm j>\vert s}\Gamma_{\rm j<\vert n}\,=\,\Gamma_{\rm j<\vert s}\Gamma_{\rm j>\vert n}\,\simeq\,\Gamma_{\rm j<\vert s}$. 
From the point of view of modelling, we recall that the post-shock Lorentz factor in the source rest frame ($\Gamma_{\rm j>\vert s}$) controls the boost factor from the emission region, while the shock Lorentz factor in the shock normal frame ($\Gamma_{\rm j<\vert n}$) controls the amount of electron heating ({\it i.e.} $\gamma_{e,\,\rm min}\simeq 600\Gamma_{\rm j<\vert n}$.

As the flow passes through the recollimation shock, it is refracted by an angle $\theta_{\rm j>}-\theta_{\rm j<}$ (in the source rest frame), which can be obtained from the post-shock and pre-shock velocities. In the small-angle approximation, meaning $\vert\theta_{\rm j<}\vert,\,\vert\theta_{\rm j>}\vert,\,\vert\alpha\vert\ll1$, we obtain
\begin{equation}
    \vert\theta_{\rm j>}-\theta_{\rm j<}\vert\,\simeq\,\frac{\vert\theta_{\rm j<}-\alpha\vert}{\sqrt{3}}\,.
\end{equation}
Therefore, if $\vert\theta_{\rm j<}-\alpha\vert\sim\mathcal O\left(1/\Gamma_{\rm j<}\right)$, the deflection through the recollimation can lead to a substantial change in the Doppler beaming from pre-shock to post-shock values. We ignore this effect in our models, however.

\section{Steady emission at a recollimation shock}
\label{app:emission}
As explained in \cite{1997ApJ...484..108S}, the Doppler amplification of the co-moving luminosity takes different forms depending on the geometric situation at hand: a moving blob of matter, or a steady jet. Both are reconciled once the steady jet is interpreted as a train of blobs moving one next to the other. The difference in amplification formula is then related to the actual number of blobs that is seen by an observer at a given time, as a result of relativistic motion.

For one moving blob, the general formula is:
\begin{equation}
    F_\nu=\frac{1+z}{d_{\rm L}^2}\,\left.\frac{{\rm d}E}{{\rm d}t_{\rm obs}{\rm d}\nu{\rm d}\Omega}\right\vert_{\vert\rm s}=\frac{1+z}{d_{\rm L}^2}\,\delta^3\,\int{\rm d}V' j'_{\nu'}\,.
    \label{eq:AB1}
\end{equation}
The subscript $_{\vert\rm s}$ indicates that the quantities are evaluated in the source rest frame. Henceforth, this subscript is dropped, and quantities written in the co-moving blob or jet frame are primed. 

In the case of a steady jet, the standard formula reads
\begin{equation}
    F_\nu=\frac{1+z}{d_{\rm L}^2}\,\frac{\delta^2}{\Gamma_{\rm b}}\,\int{\rm d}V' j'_{\nu'}\,
    \label{eq:AB5}
\end{equation}
because ${\rm d}t_{\rm obs}={\rm d}t$, {\it versus} ${\rm d}t_{\rm obs}={\rm d}t\left(1-\beta_{\rm b}\mu\right)$ for the moving blob. However, describing the steady jet as a train of blobs leads to
\citet{1997ApJ...484..108S}:
\begin{equation}
    F_\nu\,=\,\frac{1+z}{d_{\rm L}^2}\,\delta^3\,N_{\rm b,\,eff}\,\int_{V'_1}{\rm d}V' j'_{\nu'}\,,
    \label{eq:AB6}
\end{equation}
where $V'_1$ denotes the co-moving volume of one blob, of co-moving size $r'$. The total volume of the emission region (in the source frame) is assumed to be composed of $N_{\rm b}$ blobs at a given time. Because the emission time of a given blob is compressed by $(1-\beta_{\rm b}\mu)=1/\left(\Gamma_{\rm b}\delta\right)$ in the observer direction, the effective number of blobs that the observer sees at any given time is not $N_{\rm b}$, but only $N_{\rm b,\,eff}=N_{\rm b}/\left(\Gamma_{\rm b}\delta\right)$. Then both descriptions indeed match one another, given that $V=V'/\Gamma_{\rm b}$.

The total number $N_{\rm b}$ is fixed by the length scale over which emission lasts (in the source rest frame) for a given blob, that is, $d_\parallel$ (the length of the train of blobs in the source rest frame).  The co-moving duration over which a blob is active is $\delta t'=d_\parallel/(\Gamma_{\rm b}\beta_{\rm b})$ because $\delta t'=\delta t/\Gamma_{\rm b}$ and $d_\parallel = \beta_{\rm b}\delta t$ ($\delta t$ denoting the emission time in the source rest frame). The apparent width of a blob is $\delta d_\parallel = r'/\Gamma_{\rm b}$, hence
\begin{equation}
    N_{\rm b}=\frac{d_\parallel}{\delta d_\parallel}\simeq \Gamma_{\rm b}^2\beta_{\rm b}\frac{\delta t'}{r'}\,,
    \label{eq:AB7}
\end{equation}
and consequently $N_{\rm b,\,eff} \sim \delta t'/r'$. If $\delta t'\sim r'$, as in the usual blob model, then $N_{\rm b,\, eff}\sim 1$, hence the usual blob amplification formula can be used to describe the emission of the steady jet. The light curve will be roughly constant, because once a given blob has gone out of the (time) field of view of the observer, another has set in.
 
In principle, $d_\parallel$ can be fixed by other considerations than the assumption $\delta t'\sim r'$. Consider for instance the following situation, which may apply to the case of a standing shock in the flow. One may consider that the magnetic field maintains a relatively constant amplitude in a region of extent $d_\parallel$, but then decays further on due to expansion, say as a some power law of $d$. If the magnetic energy density decays faster than $d^{-1}$, then it can be shown that particles do not actually cool in the decaying part, hence the emission is truly concentrated in the region in which the magnetic field strength is roughly constant. However, if $r'$ is set by the transverse size of the structure of the recollimation shock, $d_\parallel \sim r'/\Theta_{\rm j}$ and $\Theta_{\rm j}\sim1/\Gamma_{\rm b}$, one recovers $N_{\rm b,\,eff}\sim 1$. 

Finally, one cannot exclude altogether that $N_{\rm b,\,eff}\,\sim\,{\rm a\,few}$, meaning that the observed flux would be the conjugation of several blobs at the same time. Nevertheless, adding the emission of several blobs
would be similar to assuming an elongated emission region with a smaller radius, which would not fundamentally change the resulting emission.

\section{Supplementary figures and parameters of the shock emission model}
\label{app:suppfig}

\begin{figure}[h]
   \centering
   \includegraphics[width=\hsize]{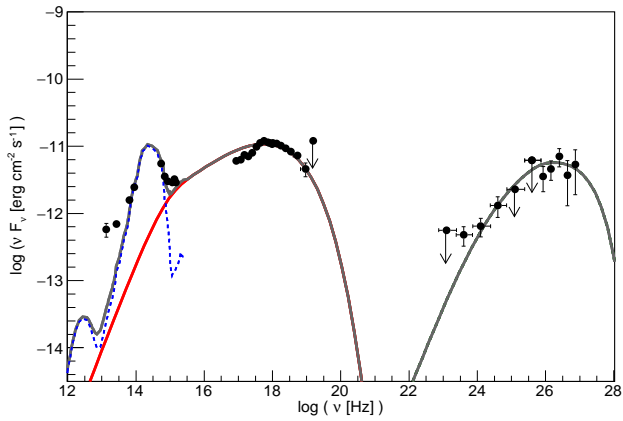}
      \caption{SSC model for the SED of RGB\,J0710+591 in scenario I. }
         \label{fig:sed0710}
   \end{figure}

\begin{figure}[h]
   \centering
   \includegraphics[width=\hsize]{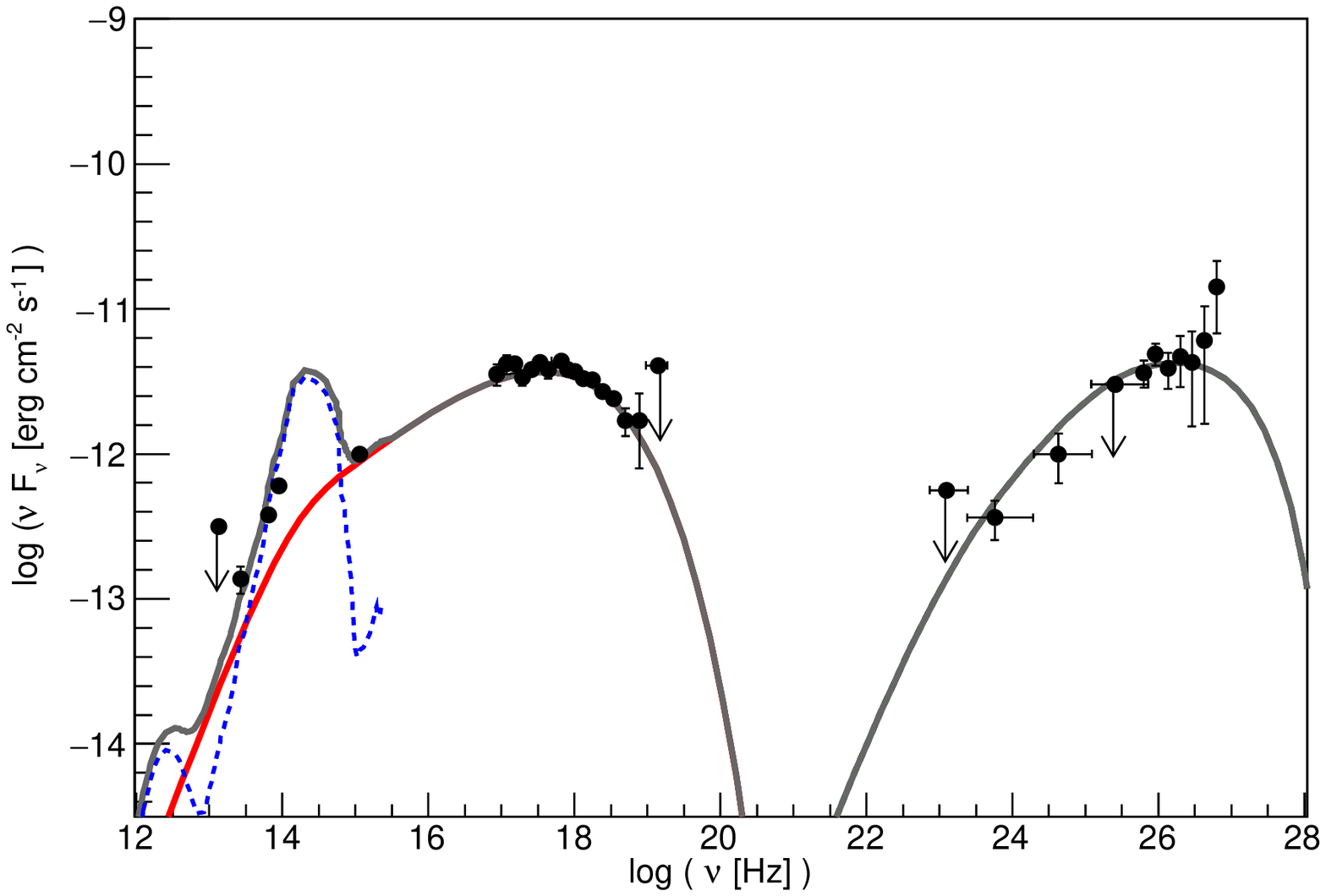}
      \caption{SSC model for the SED of 1ES\,0347-121 in scenario I. }
         \label{fig:sed0347}
   \end{figure}

\begin{figure}[h]
   \centering
   \includegraphics[width=\hsize]{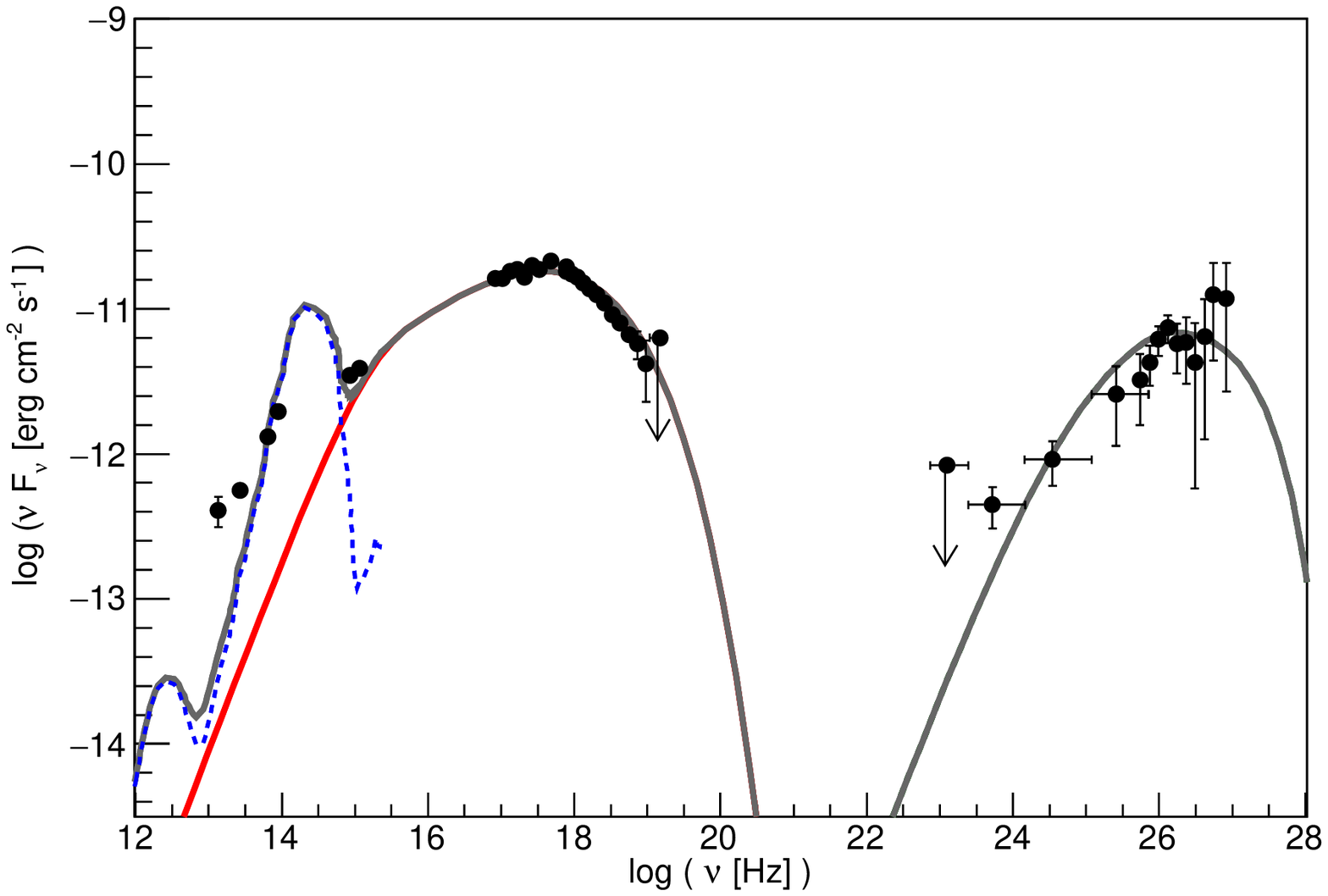}
      \caption{SSC model for the SED of 1ES\,1101-232 in scenario I. }
         \label{fig:sed1101}
   \end{figure}

\begin{figure}[h]
   \centering
   \includegraphics[width=\hsize]{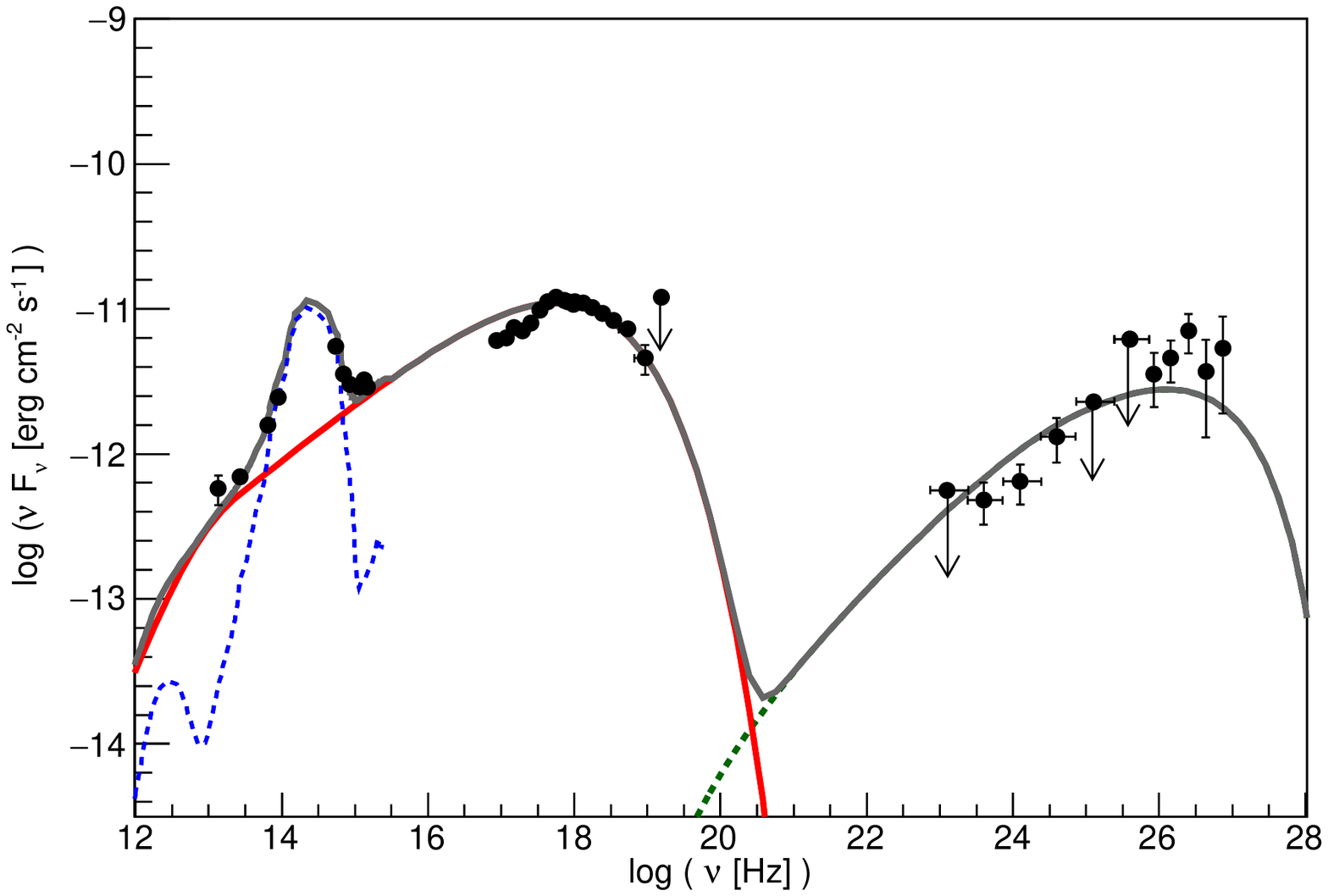}
      \caption{SSC model for the SED of RGB\,J0710+591 in scenario II. }
         \label{fig:sed0710_II}
   \end{figure}

\begin{figure}[h]
   \centering
   \includegraphics[width=\hsize]{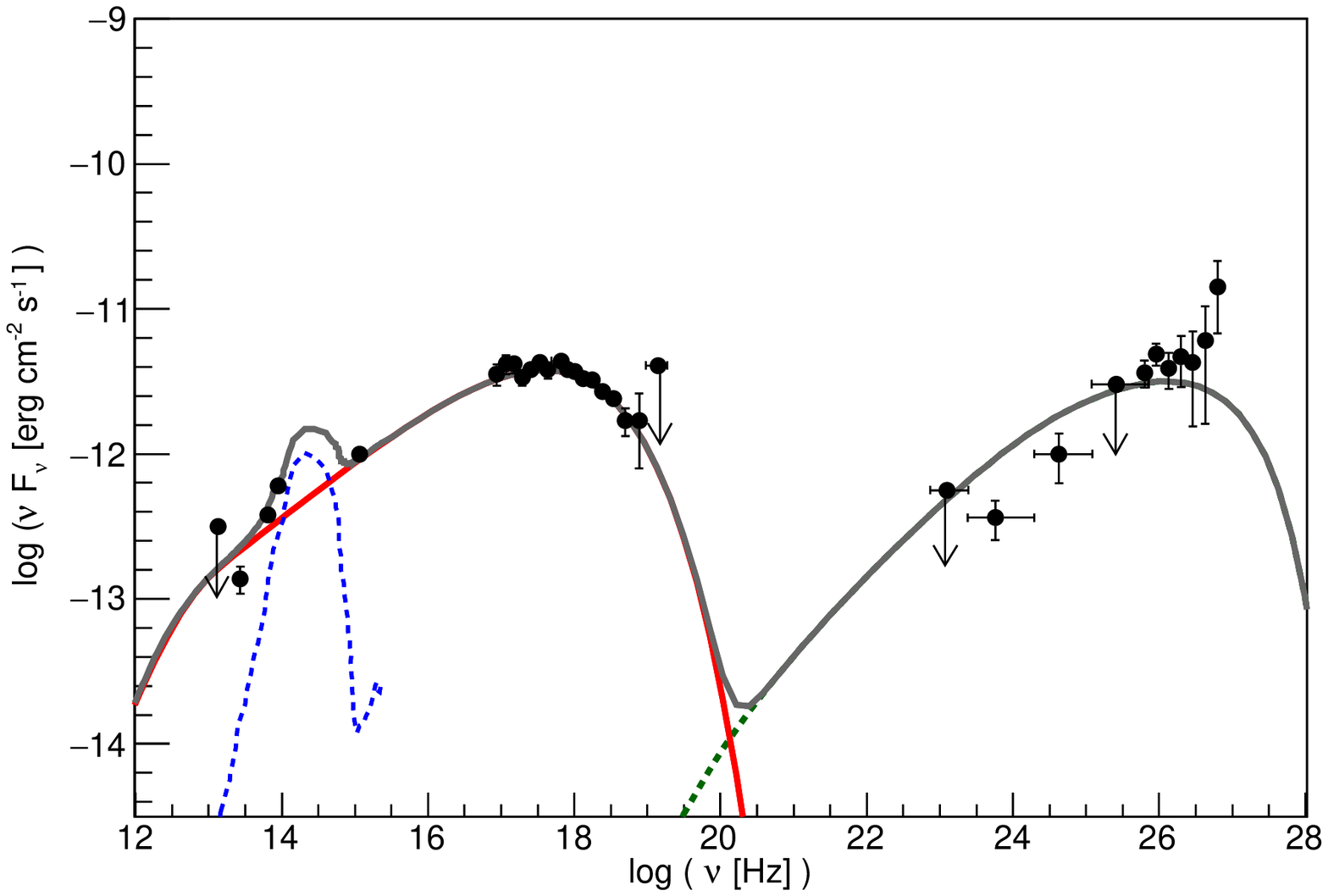}
      \caption{SSC model for the SED of 1ES\,0347-121 in scenario II. }
         \label{fig:sed0347_II}
   \end{figure}

\begin{figure}[h]
   \centering
   \includegraphics[width=\hsize]{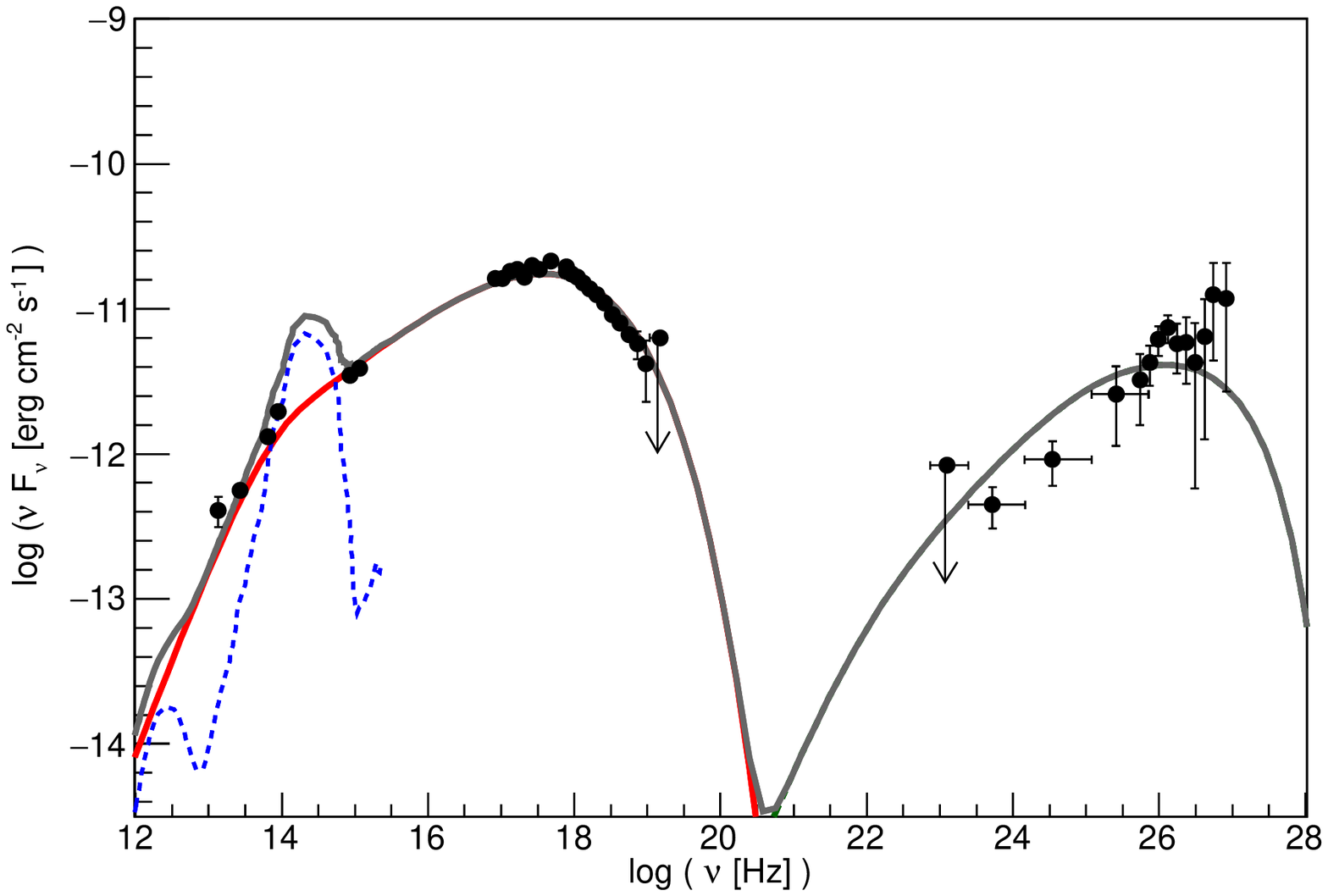}
      \caption{SSC model for the SED of 1ES\,1101-232 in scenario II. }
         \label{fig:sed1101_II}
   \end{figure}

\begin{figure}[h]
   \centering
   \includegraphics[width=\hsize]{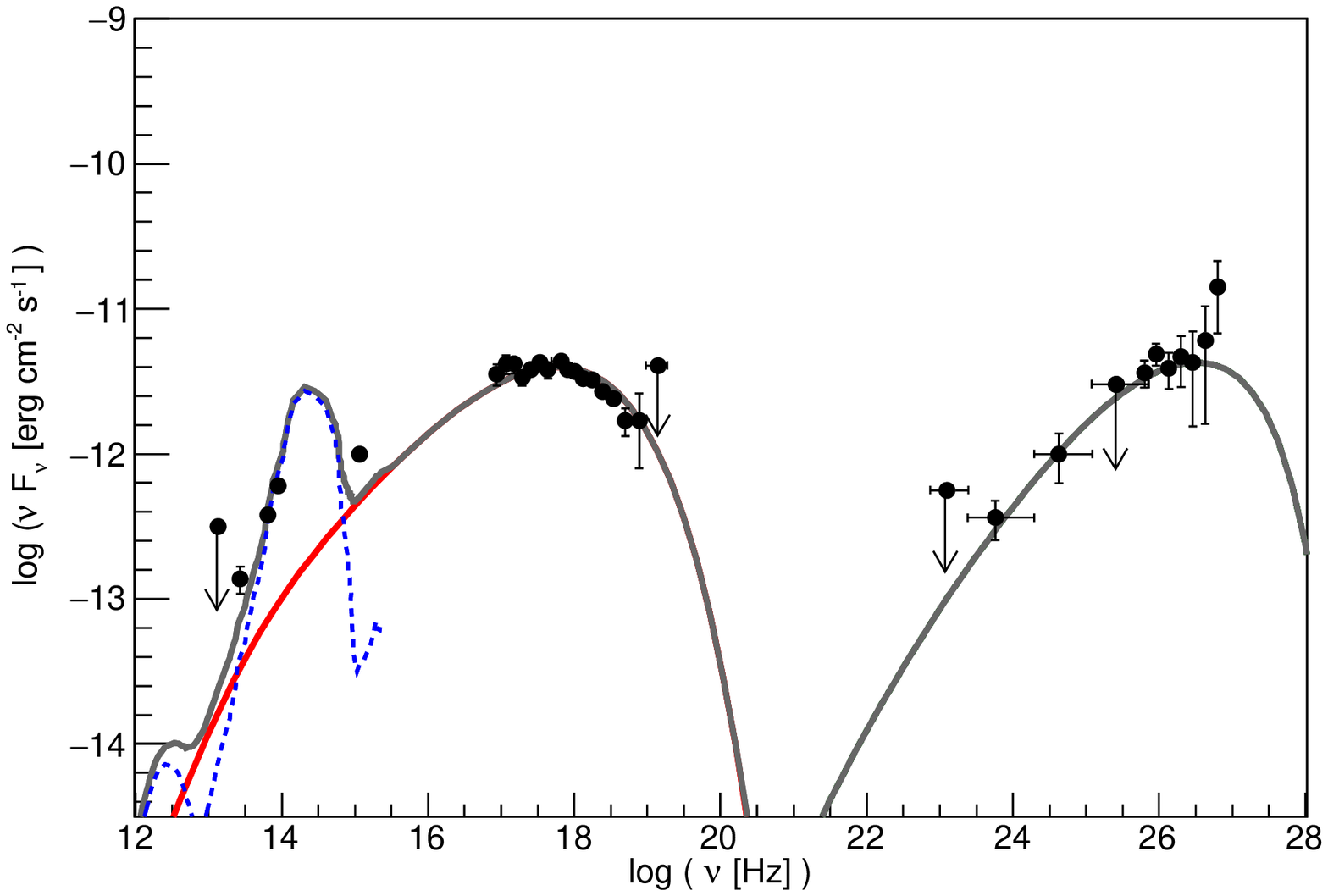}
      \caption{SSC model for the SED of 1ES\,0347-121 in scenario III.}
         \label{fig:sed0347_III}
   \end{figure}

Solutions for the SEDs of the sources RGB\,J0710+591, 1ES\,0347-121 and 1ES\,1101-232 for the generic model (scenario I) are
shown in Figs.~\ref{fig:sed0710} to~\ref{fig:sed1101}. For the same sources, solutions of the recollimation-shock model (scenario II) are shown in Figs.~\ref{fig:sed0710_II} to~\ref{fig:sed1101_II}. Figures~\ref{fig:sed0347_III} and~\ref{fig:sed1101_III} show the
solutions with a particle population that has been re-accelerated on a second shock (scenario III) for the sources 1ES\,0347-121 and 1ES\,1101-232, which feature very hard spectra in the {\it Fermi}-LAT range.

Observables used as an input in the search for model solutions in all three scenarios, as well as parameters of the proton populations, are shown in Table~\ref{tab:par2}.

\begin{figure}[!h]
   \centering
   \includegraphics[width=\hsize]{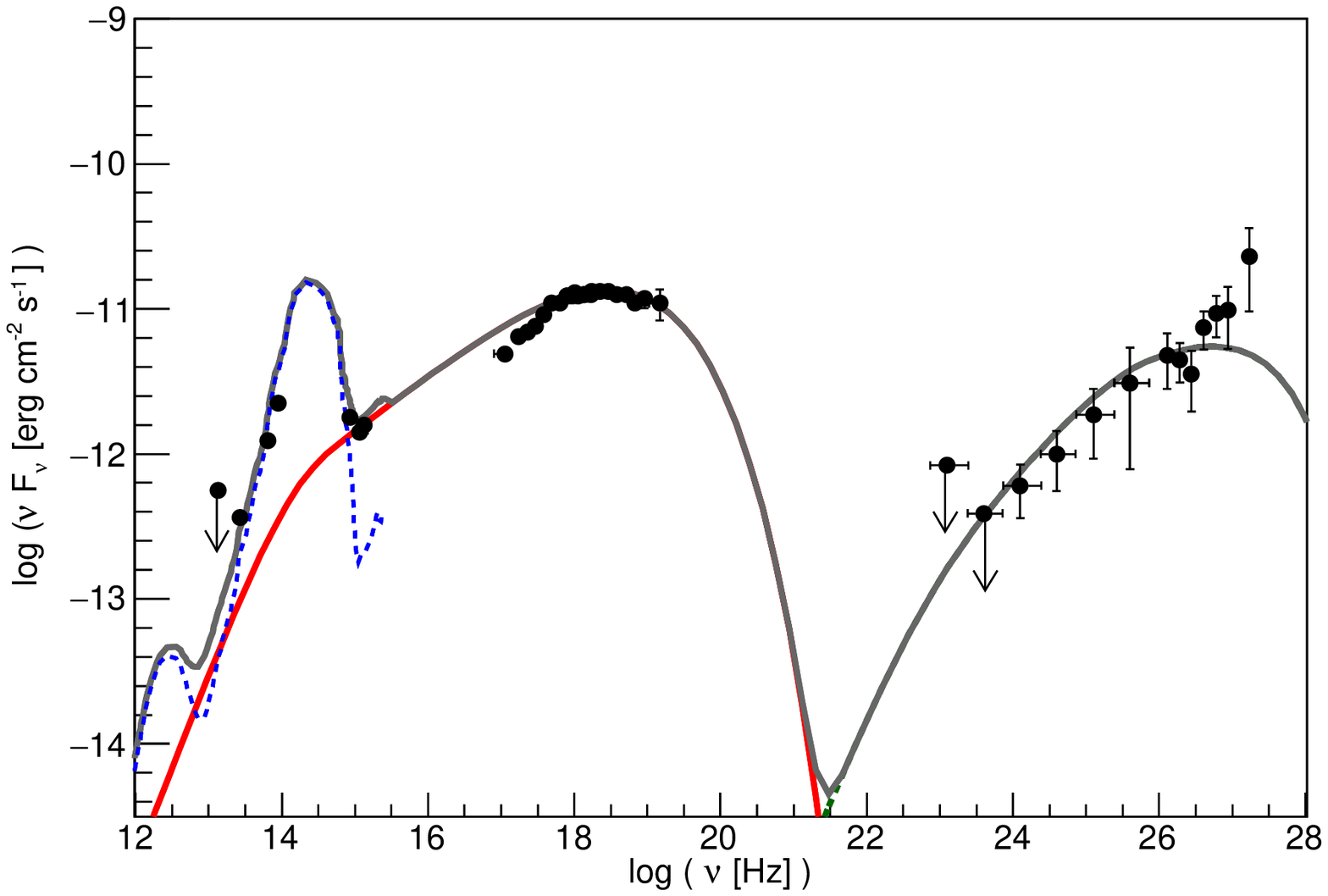}
      \caption{SSC model for the SED of 1ES\,0229+200 in scenario II. }
         \label{fig:sed0229_II}
   \end{figure}

        \begin{table*}
   \centering
   \caption{Observables and proton parameters used for the modelling of the extreme blazar datasets in scenarios I, II, and III.}
                \begin{tabular}{l c c c c c  }
                \hline
                \hline
                 \noalign{\smallskip}
                &1ES 0229+200 &1ES 0347-121  &RGB J0710+591 & 1ES 1101-232 & 1ES 1218+304\\
                 \noalign{\smallskip}
                \hline
            z & 0.140 & 0.188 & 0.125 & 0.186 & 0.184  \\
            $L_{\rm Edd} \, [10^{46} {\rm erg \, s}^{-1}]$ & 18 & 1.3 & 2.2 & 13 & 1.4 \\
            $t_{\rm var, obs} \, [{\rm s}]$ & $\sim 10^7$ & - & - & - & $ \sim 10^5$\\
            $\nu_{\rm syn} \, [{\rm Hz}] $& $3 \times 10^{18} $ & $ 4 \times 10^{17}$ & $6 \times 10^{17}$ & $ 2-4 \times 10^{17}$ & $2 \times 10^{17}$ \\
            $\nu_{\rm IC} \, [{\rm Hz}] $& $1 \times 10^{27}$ & $3 \times 10^{26}$ & $3 \times 10^{26}$ & $2.5 \times 10^{26}$& $2 \times 10^{26}$ \\
           $ \nu F_{\nu}(\nu_{\rm syn}) \, [{\rm erg \, cm^{-2} \, s}^{-1}] $ & $ 2.0 \times 10^{-11}$ & $ 5.0 \times 10^{-12} $ & $1.3 \times 10^{-11}$ & $2.0 \times 10^{-11}$ & $ 1.0 \times 10^{-11}$ \\
           $ \nu F_{\nu}(\nu_{\rm IC}) \, [{\rm erg \, cm^{-2} \, s}^{-1}] $ & $5.0 \times 10^{-12}$ & $ 5.0 \times 10^{-12} $ & $6.3 \times 10^{-12}$ & $6.3 \times 10^{-12}$ & $1.5 \times 10^{-11}$\\
            \noalign{\smallskip}
                \hline
                 \noalign{\smallskip}
         scenario I & & & & &\\
            \noalign{\smallskip}
                \hline
                 \noalign{\smallskip}
                $\gamma_{p,\rm min}$& 50 & 30 & 40 & 50 & 8 \\
                $\gamma_{p,\rm max}$& $4.9 \times 10^{3}$ & $2.5 \times 10^3$ & $1.8 \times 10^{3}$ & $1.3 \times 10^{3}$ & $840$  \\
                $L_{\rm jet} \, [10^{44} \, {\rm erg \, s}^{-1}]$ & 3.6 & 5.1 & 2.1 & 2.9 & 8.3 \\
                \noalign{\smallskip}
                \hline
                 \noalign{\smallskip}
                 scenario II & & & & & \\
                 \noalign{\smallskip}
                \hline
                 \noalign{\smallskip}
                $\gamma_{p,\rm min}$& 20 & 3 & 3 & 10 & 3  \\
                $\gamma_{p,\rm max}$& $1.5 \times 10^{3}$ &  280 & 87 & 160 & 250 \\
                $L_{\rm jet} \, [10^{44} \, {\rm erg \, s}^{-1}]$ & 3.6 & 5.7 & 2.1 & 2.9 & 9.2 \\
                \noalign{\smallskip}
                \hline
                 \noalign{\smallskip}           
                                 scenario III & & & & & \\
                 \noalign{\smallskip}
                \hline
                 \noalign{\smallskip}
                $\gamma_{p,\rm min}$& 3.0 & 3.0 & - & 3.0 & - \\
                $\gamma_{p,\rm max}$& 300. & 320. & - & 47. & - \\
                $L_{\rm jet} \, [10^{44} \, {\rm erg \, s}^{-1}]$  & 3.7 & 2.6 & - & 2.5 & - \\
                \noalign{\smallskip}
                \hline
                \hline          
                 \noalign{\bigskip}     
                \end{tabular}
         \label{tab:par2}
         \newline
                \end{table*}

\end{appendix}

\end{document}